\documentclass[aps,preprintnumbers]{revtex4}
\usepackage{amsmath}
\usepackage{graphicx}

\input{tcilatex}

\begin{document}
\title{Atom interferometer as a selective sensor of rotation or gravity}
\author{B. Dubetsky and M. A. Kasevich}
\affiliation{Department of Physics,Stanford University,Stanford, California 94305-4060,USA.}
\date{10 April 2006}

\begin{abstract}
In the presence of Earth gravity and gravity-gradient forces, centrifugal and
Coriolis forces caused by the Earth rotation, the phase of the time-domain
atom interferometers is calculated with accuracy up to the terms proportional
to the fourth degree of the time separation between pulses. We considered
double-loop atom interferometers and found appropriate condition to eliminate
their sensitivity to acceleration to get atomic gyroscope, or to eliminate the
sensitivity to rotation to increase accuracy of the atomic gravimeter.
Consequent use of these interferometers allows one to measure all components
of the acceleration and rotation frequency projection on the plane
perpendicular to gravity acceleration. Atom interference on the Raman
transition driving by noncounterpropagating optical fields is proposed to
exclude stimulated echo processes which can affect the accuracy of the atomic
gyroscopes. Using noncounterpropagating optical fields allows one to get a new
type of the Ramsey fringes arising in the unidirectional Raman pulses and
therefore centered at the two-quantum line center. Density matrix in the
Wigner representation is used to perform calculations. It is shown that in the
time between pulses, in the noninertial frame, for atoms with fully quantized
spatial degrees of freedom, this density matrix obeys classical Liouville equations.

\end{abstract}

\maketitle

\section{\label{s1}Introduction}

Since atom interference \cite{c1} has been proposed as a sensor of inertial
effects \cite{c2} and the use of Raman transition between atomic hyperfine
sublevels \cite{c3} allowed tremendously increase the time separation between
optical fields, unprecedented accuracy in the measurement of the Earth
rotation \cite{c4}, gravity gradients \cite{c5}, $\hbar/m_{Cs}$ \cite{c6}, and
accelerations \cite{c7} has also been achieved. The theoretical analysis
\cite{c71,c72,c8,c9} showed that the current level of interferometer phase
measurements is sufficient to sense each source changing the atomic motion,
i.e., rotation, acceleration, acceleration gradient, and recoil effect
\cite{c10}, as well as the interplay between them, such as between rotation
and acceleration \cite{c8}, recoil effect and rotation \cite{c11}, recoil
effect and gravity gradient\cite{c8}.

At the same time for navigation and geodetic applications one needs to measure
separately a gravity acceleration $\mathbf{g}$ and rotation frequency
$\mathbf{\Omega},$ i.e., one needs an interferometer whose phase is
selectively sensitive only to one component of these vectors, while the
sensitivity to others can be either excluded with sufficiently high accuracy
or precisely taken into account. An example here is an atomic gyro \cite{c4},
where the $\mathbf{\Omega}$ projection on the gravity acceleration
$\mathbf{g}$ has been measured, because influence of gravity has been excluded
using spatially separated fields propagating in horizontal plane and using
signals from two counterpropagating atomic beams.

In this article we consider the theory of atom interferometers that would
allow one to measure separately an atom acceleration $\mathbf{g}$ and the
rotation frequency component $\mathbf{\Omega}$ which are perpendicular to
$\mathbf{g}.$ We explore the fact that, when rotation, gravity gradient and
recoil effect only slightly affect the atoms' trajectory, parts of the
interferometer phase associated with $\mathbf{g}$ and $\mathbf{\Omega}$ evolve
correspondingly as $T^{2}$ and $T^{3}$ while other contributions are precisely
known or negligibly small. Using particular kinds of the atom interferometers
one can exclude one of these dependences getting a sensor of acceleration or
rotation only.

\begin{figure} [htbp]
\vspace{0.5cm}
\includegraphics[height=12.0cm,width=12.0cm]{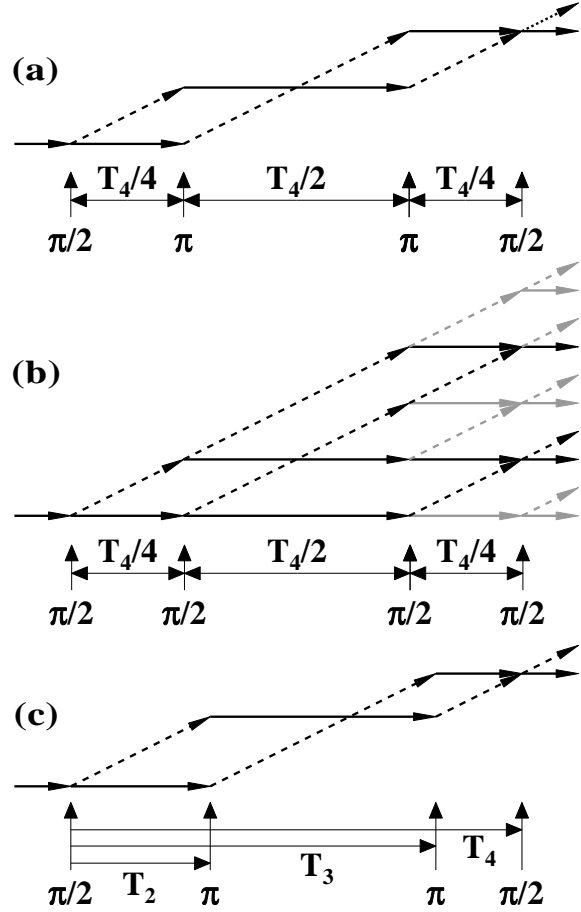}
\caption{Recoil diagrams for four-pulse atom interferometers, $\left(
a\right)  $ double-loop interferometer insensitive to homogeneous
acceleration; $\left(  b\right)  $ stimulated echo processes, shown by black
parallelograms, which could restore sensitivity to acceleration; $\left(
c\right)  $ double-loop interferometer with nonidentical loops, $T_{2,3}%
=\dfrac{\sqrt{5}\mp1}{4}T_{4}.$ Solid and dashed lines correspond to the lower
and upper atomic states.}
\label{f1}
\end{figure}

Usual time domain atom interferometer, consisting of three resonant pulses
applied to the cold atom cloud at moments $t=0,$ $T_{2}$ and $T_{3}$ could not
serve for these purposes because the only parameter, ratio of the time
separation between pulses $T_{3}/T_{2}$, is already used to eliminate linear
in time part of the phase. This Doppler phase vanishes at the echo point,
$T_{3}=2T_{2}.$ To eliminate another terms one needs at least a four-pulse
interferometer. It was found previously \cite{c2,c121,c13} that double loop
interferometer consisting of two $\pi/2$ pulses at the beginning and end and
two $\pi$ pulses in between, i.e., the $\left(  \pi/2,T_{1}=0\right)  -\left(
\pi,T_{2}\right)  -\left(  \pi,T_{3}\right)  -\left(  \pi/2,T_{4}\right)  $
interferometer, has no sensitivity to the homogeneous acceleration if
$T_{2}=\dfrac{1}{4}T_{4}$ and $T_{3}=\dfrac{3}{4}T_{4},$ see fig. \ref{f1}(a).

One can expect that the double-loop interferometer can serve as a sensor of
rotation. It is correct only if pulses' areas are precisely equal to assigned
values. If one could not hold areas equal to $\pi$ or $\pi/2,$ then stimulated
echo processes also contribute [see Fig. \ref{f1}(b)] to the interferometer
phase and since stimulated echo is insensitive to the atomic motion between
second and third pulses, their contribution is acceleration sensitive. Since
typically contribution to the interferometers phase caused by the acceleration
is 5-6 orders of magnitude larger than that caused by rotation \cite{c121}, it
is extremely important to find a technique for excluding stimulated echo. To
achieve this goal one should violate the condition for stimulated echo. This
opportunity exists if one uses Raman transitions, where the splitting of
atomic momentum states \ occurs under the action of two optical fields having
wave vectors $\mathbf{q}_{1}$ and $\mathbf{q}_{2},$ such as an effective wave
vector associated with Raman transition is given by $\mathbf{k}=\mathbf{q}%
_{1}-\mathbf{q}_{2}.$ In principle one can use for this purpose
noncouterpropagating fields. Interferometers with $\mathbf{q}_{1}%
\not =-\mathbf{q}_{2}$ have been previously created \cite{c131} and studied
\cite{c132}.

Since for Raman transition between hyperfine sublevels $q_{1}\approx q_{2},$
$\left\vert \mathbf{k}\right\vert \approx k\sin\alpha,$ where $\alpha$ is the
half-angle between vectors $\mathbf{q}_{1}$ and $\mathbf{q}_{2},$ and
$k\approx2q_{1}$ is the maximum value of the effective wave vector
corresponding to counterpropagating optical fields. If $\mathbf{k}_{j}$ is an
effective wave vector associated with Raman pulse $j,$ than the phase matching
condition for pulse sequences $\pi/2-\pi-\pi-\pi/2$ [Fig. \ref{f1}(a)] and
$\pi/2-\pi/2-\pi/2-\pi/2$ [Fig. \ref{f1}(b)] are%
\begin{subequations}%
\label{1}%
\begin{align}
&  \mathbf{k}_{4}-2\mathbf{k}_{3}+2\mathbf{k}_{2}-\mathbf{k}_{1}%
=0,\label{1a}\\
&  \mathbf{k}_{4}-\mathbf{k}_{3}-\mathbf{k}_{2}+\mathbf{k}_{1}=0.\label{1b}%
\end{align}%
\end{subequations}%
These conditions mean the same only for parallel pulses having effective wave
vectors of the same absolute values $k_{j}$. But even for pulses with slightly
different $k_{j}$ one can hold only one of this condition and violate another.
In this paper we propose and analyze an exclusion of the stimulated echo
processes using for double loop atom interferometers Raman pulses with
nonequal effective wave vectors.

While eliminating $T^{2}$ dependences one can use time domain interferometer
as a gyroscope, to measure rotation frequency components perpendicular to
acceleration, eliminating $T^{3}$ dependences would guarantee that
contributions of this order, which can be caused by \cite{c8} gravity gradient
(for atoms with nonzero launch momentum), rotation (when effective wave
\ vectors are not parallel to the acceleration), recoil effect and
combinations of these factors, do not affect the interferometer phase at all.
Absence of $T^{3}$ dependences should increase the accuracy of the
acceleration measurement. We will show below that $T^{3}$ dependences
disappear for the double-loop interferometer with nonidentical loops shown in
Fig. \ref{f1}(c). Since in this case pulse separations $T_{2},$ $T_{3}$ and
$T_{4}$ are incommensurate, one can be sure that no other echolike processes
can affect the interferometer phase.

In this paper we perform calculations of interferometers' phases for three-
and four-pulse cases, for arbitrarily directed gravity acceleration and
rotation including Earth gravity gradient as a small correction. It would
allow us to get{} the phases' terms evolving up to $T^{4}$ dependences. In our
calculation we use equations for the density matrix in the Wigner
representation, which are convenient to describe atomic clouds' evolution
between pulses \cite{c14}, because we show that in this representation density
matrix evolves exactly as the classical distribution function in the rotating
frame in the presence of the homogeneous gravity and gravity gradient terms.
This technique is mathematically {}equivalent to the pure quantum
consideration of the atomic wave function evolution in the space between
pulses elaborated in \cite{c72}.

Using density matrix in the Wigner representation allows us to evaluate
precisely all effects related to the atomic spatial motion quantization
without calculating path integrals \cite{c15}.

The paper is arranged as follows. A simplified consideration of four-pulse
interferometer is performed in Sec. \ref{s2}. Evolution of the density matrix
in the Wigner representation is considered in Sec. \ref{s3}. Atom trajectory
in the Earth rotating frame, including gravity-gradient terms as perturbation,
is calculated in Sec. \ref{s4}. Density matrix jumps under the pulse action
are obtained in Sec. \ref{s5}. In Sec. \ref{s6} we calculate the phase of the
$\pi/2-\pi-\pi/2$ pulse interferometer. Section \ref{s7} is devoted to study
four-pulse interferometers as atomic gyroscope or gravimeter, while in Sec.
\ref{s8} we summarize our consideration and discuss further possible
developments of the proposed multiple loop interferometers technique.

\section{\label{s2}Four-pulse atom interferometers}

Consider an interaction of cold atoms with a sequence of four Raman pulses,
$\pi/2-\pi-\pi-\pi/2,$ having the same effective wave vector $\mathbf{k}$ and
applied at moments $T_{1}=0,$ $T_{2}$, $T_{3},$ and $T_{4}.$ We will show
below that for atoms with initial coordinate and momentum $\mathbf{r}$ and
$\mathbf{p}$ the phase of interferometer associated with atomic motion is
given by%
\begin{equation}
\phi_{m}=\mathbf{k}\cdot\left[  \mathbf{R}\left(  \mathbf{r},\mathbf{p}%
+\dfrac{\hbar\mathbf{k}}{2},T_{4}\right)  -2\mathbf{R}\left(  \mathbf{r}%
,\mathbf{p}+\dfrac{\hbar\mathbf{k}}{2},T_{3}\right)  +2\mathbf{R}\left(
\mathbf{r},\mathbf{p}+\dfrac{\hbar\mathbf{k}}{2},T_{2}\right)  -\mathbf{r}%
\right]  ,\label{2}%
\end{equation}
where the function $\mathbf{R}\left(  \mathbf{r},\mathbf{p},t\right)  $ is the
position of an atom at the moment $t$ under the classical motion in rotating
frame. For small rotation frequency $\mathbf{\Omega}$ $\left(  \Omega
t\ll1\right)  $ the atom trajectory is given by
\begin{equation}
\mathbf{R}\left(  \mathbf{r},\mathbf{p},t\right)  \approx\mathbf{r}%
+\dfrac{\mathbf{p}}{m}t+\mathbf{g}\dfrac{t^{2}}{2}-\mathbf{\Omega}%
\times\left(  \mathbf{r}+\dfrac{\mathbf{p}}{m}t+\mathbf{g}\dfrac{t^{2}}%
{3}\right)  t,\label{3}%
\end{equation}
where $m$ is the atomic mass, and for the phase (\ref{2}) one arrives at the
expression%
\begin{align}
\phi_{m} &  =\mathbf{k}\cdot\left(  \dfrac{\mathbf{p}+\dfrac{\hbar\mathbf{k}%
}{2}}{m}-\mathbf{\Omega}\times\mathbf{r}\right)  T_{4}\left(  1-2t_{3}%
+2t_{2}\right)  \nonumber\\
&  +\mathbf{k}\cdot\left(  \dfrac{\mathbf{g}}{2}-\mathbf{\Omega}\times
\dfrac{\mathbf{p}}{m}\right)  T_{4}^{2}\left(  1-2t_{3}^{2}+2t_{2}^{2}\right)
\nonumber\\
&  +\dfrac{1}{3}\left(  \mathbf{k}\times\mathbf{g}\right)  \cdot
\mathbf{\Omega}T_{4}^{3}\left(  1-2t_{3}^{3}+2t_{2}^{3}\right)  ,\label{4}%
\end{align}
where%
\begin{equation}
t_{j}=T_{j}/T_{4}.\label{401}%
\end{equation}
If the purpose is to measure the rotation frequency, then one may set to $0$
first and second terms in Eq. (\ref{4}). System%
\begin{subequations}%
\label{5}%
\begin{align}
1-2t_{3}+2t_{2} &  =0,\label{5a}\\
1-2t_{3}^{2}+2t_{2}^{2} &  =0\label{5b}%
\end{align}%
\end{subequations}%
has a solution $t_{2}=1/4,t_{3}=3/4$ corresponding to the double loop
interferometer \cite{c2,c121,c13} shown in Fig. \ref{f1}(a). In this case
\begin{equation}
\phi_{m}=\dfrac{1}{16}\left(  \mathbf{k}\times\mathbf{g}\right)
\cdot\mathbf{\Omega}T_{4}^{3}.\label{6}%
\end{equation}
If the gravity is measured, then the phase (\ref{6}) can be used to measure
perpendicular to gravity components of the rotation frequency. For example,
for Cs interferometer having the effective wavelength $426~$nm and total time
separation between pulses $T_{4}=0.8$ s, at the latitude 41$^{\circ}$ the
maximum value of the phase (\ref{6}) $\phi_{m}\approx250.$

If the purpose is to eliminate cubic terms in order to increase the accuracy
of the acceleration measurement, one should choose the solution of the system%
\begin{subequations}%
\label{7}%
%

\begin{align}
1-2t_{3}+2t_{2} &  =0,\label{7a}\\
1-2t_{3}^{3}+2t_{2}^{3} &  =0,\label{7b}%
\end{align}%
\end{subequations}%
which is
\begin{equation}
t_{2,3}=\dfrac{\sqrt{5}\mp1}{4},\label{8}%
\end{equation}
and for atoms having zero launch momentum, $\mathbf{p=0,}$ the phase is given
by%
\begin{equation}
\phi_{m}=-\dfrac{\sqrt{5}-2}{4}\mathbf{k}\cdot\mathbf{g}T_{4}^{2}.\label{9}%
\end{equation}
For a given maximum separation between pulses, the absolute value of the phase
(\ref{9}) is approximately $4.2$ times less than the signal in the usual
three-pulse interferometer. Therefore, it is a kind of trade off, when one
should decide whether it is more important to get a larger signal or to
eliminate $T^{3}$ corrections to the phase.

\section{\label{s3}Density matrix evolution in the free space}

Starting with the Lagrangian for an atom in the frame rotating with constant
rate,%
\begin{equation}
L=\dfrac{m\mathbf{v}^{2}}{2}+m\mathbf{v}\cdot\left(  \mathbf{\Omega}%
\times\mathbf{r}\right)  +\dfrac{m}{2}\left(  \mathbf{\Omega}\times
\mathbf{r}\right)  ^{2}-U,\label{10}%
\end{equation}
where $U$ is the potential, one gets the generalized momentum and Hamiltonian,%
\begin{subequations}%
\label{11}
\begin{align}
\mathbf{p} &  =m\mathbf{v}+m\left(  \mathbf{\Omega}\times\mathbf{r}\right)
;\label{11a}\\
H &  =\dfrac{\mathbf{p}^{2}}{2m}+\mathbf{p\cdot}\left(  \mathbf{r}%
\times\mathbf{\Omega}\right)  +U.\label{11b}%
\end{align}%
\end{subequations}%
In the space free of the laser field, i.e., free space, the density matrix in
the Wigner representation \cite{c151}
\begin{equation}
\rho\left(  \mathbf{r},\mathbf{p},t\right)  =\dfrac{1}{\left(  2\pi
\hbar\right)  ^{3}}\int d\mathbf{s}\rho\left(  \mathbf{r}+\dfrac{1}%
{2}\mathbf{s},\mathbf{r}-\dfrac{1}{2}\mathbf{s},t\right)  \exp\left(
-i\mathbf{p}\cdot\mathbf{s}/\hbar\right)  ,\label{12}%
\end{equation}
where%
\begin{equation}
\rho\left(  \mathbf{r},\mathbf{r}^{\prime}\right)  =\psi\left(  \mathbf{r}%
\right)  \otimes\psi^{\dag}\left(  \mathbf{r}^{\prime}\right)  \label{13}%
\end{equation}
is the density matrix in the coordinate representation, evolves as%
\begin{align}
\partial_{t}\rho\left(  \mathbf{r},\mathbf{p},t\right)   &  =\dfrac{1}%
{i\hbar\left(  2\pi\hbar\right)  ^{3}}\int d\mathbf{s}\exp\left(
-i\mathbf{p}\cdot\mathbf{s}/\hbar\right)  \left\{  \left[  -\dfrac{\hbar^{2}%
}{2m}\partial_{\mathbf{r}_{1}}^{2}+i\hbar\partial_{\mathbf{r}_{1}}\left(
\mathbf{\Omega}\times\mathbf{r}_{1}\right)  +U\left(  \mathbf{r}_{1}\right)
\right.  \right.  \nonumber\\
&  +\left.  \left.  \dfrac{\hbar^{2}}{2m}\partial_{\mathbf{r}_{2}}^{2}%
+i\hbar\partial_{\mathbf{r}_{2}}\left(  \mathbf{\Omega}\times\mathbf{r}%
_{2}\right)  -U\left(  \mathbf{r}_{2}\right)  \right]  \rho\left(
\mathbf{r}_{1},\mathbf{r}_{2},t\right)  \right\}  _{\mathbf{r}_{1}%
=\mathbf{r}+\frac{1}{2}\mathbf{s},\mathbf{r}_{2}=\mathbf{r}-\frac{1}%
{2}\mathbf{s}}.\label{14}%
\end{align}
Using the fact that
\begin{align}
\partial_{\mathbf{r}_{1}} &  =\dfrac{1}{2}\partial_{\mathbf{r}}+\partial
_{\mathbf{s}};\nonumber\\
\partial_{\mathbf{r}_{2}} &  =\dfrac{1}{2}\partial_{\mathbf{r}}-\partial
_{\mathbf{s}}\label{15}%
\end{align}
and replacing $\partial_{\mathbf{s}}\Rightarrow i\mathbf{p}/\hbar,$ one
arrives at equations%
\begin{subequations}%
\label{151}
\begin{gather}
\left\{  \partial_{t}+\left(  \dfrac{\mathbf{p}}{m}+\mathbf{r}\times
\Omega\right)  \partial_{\mathbf{r}}+\left[  \left(  \mathbf{p}\times
\Omega\right)  -\partial_{\mathbf{r}}U\right]  \partial_{\mathbf{p}%
}+Q\right\}  \rho\left(  \mathbf{r},\mathbf{p},t\right)  =0;\label{151a}\\
Q=-\left(  i\hbar\right)  ^{-1}\left[  U\left(  \mathbf{r}+\dfrac{1}{2}%
i\hbar\partial_{\mathbf{p}}\right)  -U\left(  \mathbf{r}-\dfrac{1}{2}%
i\hbar\partial_{\mathbf{p}}\right)  \right]  +\partial_{\mathbf{r}}%
U\partial_{\mathbf{p}}.\label{151b}%
\end{gather}%
\end{subequations}%
When the potential $U$ is not higher than bilinear function of coordinates,
quantum term disappears,\textbf{ }$Q=0.$ Since the Hamiltonian equation for
the classical momentum and coordinate is
\begin{subequations}%
\label{16}%
\begin{align}
\mathbf{\dot{r}} &  =\dfrac{\mathbf{p}}{m}+\mathbf{r}\times\mathbf{\Omega
};\label{16a}\\
\mathbf{\dot{p}} &  =-\partial_{\mathbf{r}}U+\left(  \mathbf{p}\times
\mathbf{\Omega}\right)  ,\label{16b}%
\end{align}%
\end{subequations}%
one concludes that, if
\begin{equation}
Q\sim\hbar^{2}\left(  \partial_{\mathbf{r}}^{3}U\right)  \partial_{\mathbf{p}%
}^{3}\ll T^{-1},\label{17}%
\end{equation}
where $T$ is a typical time of the system's evolution, then with an accuracy
$QT$ the quantum density matrix in the Wigner representation obeys the same
equation as the classical density matrix%
\begin{equation}
\left(  \partial_{t}+\mathbf{\dot{r}}\partial_{\mathbf{r}}+\mathbf{\dot{p}%
}\partial_{\mathbf{p}}\right)  \rho\left(  \mathbf{r},\mathbf{p},t\right)
=0.\label{18}%
\end{equation}

If the density matrix is known at some previous moment $t^{\prime}<t$ then the
solution of Eq. (\ref{18}) is given by%
\begin{equation}
\rho\left(  \mathbf{r},\mathbf{p},t\right)  =\rho\left[  \mathbf{R}\left(
\mathbf{r},\mathbf{p},t^{\prime}-t\right)  ,\mathbf{P}\left(  \mathbf{r}%
,\mathbf{p},t^{\prime}-t\right)  ,t^{\prime}\right]  ,\label{19}%
\end{equation}
where functions $\mathbf{R}\left(  \mathbf{r}_{0},\mathbf{p}_{0},t\right)
,\mathbf{P}\left(  \mathbf{r}_{0},\mathbf{p}_{0},t\right)  $ are solutions of
Eqs. (\ref{16}) subjected to the initial condition $\left(  \mathbf{r}%
,\mathbf{p}\right)  _{t=0}=\left(  \mathbf{r}_{0},\mathbf{p}_{0}\right)  .$
Clearly these functions have to satisfy multiplication law%
\begin{subequations}%
\label{20}%
\begin{align}
\mathbf{R}\left[  \mathbf{R}\left(  \mathbf{r},\mathbf{p},t_{1}\right)
,\mathbf{P}\left(  \mathbf{r},\mathbf{p},t_{1}\right)  ,t_{2}\right]   &
=\mathbf{R}\left(  \mathbf{r},\mathbf{p},t_{1}+t_{2}\right)  ,\label{20a}\\
\mathbf{P}\left[  \mathbf{R}\left(  \mathbf{r},\mathbf{p},t_{1}\right)
,\mathbf{P}\left(  \mathbf{r},\mathbf{p},t_{1}\right)  ,t_{2}\right]   &
=\mathbf{P}\left(  \mathbf{r},\mathbf{p},t_{1}+t_{2}\right)  .\label{20b}%
\end{align}%
\end{subequations}%
Solution (\ref{19}) is the consequence of the phase-space invariance in time.
This is correct only for canonical variables $\left(  \mathbf{r}%
,\mathbf{p}\right)  $\ and incorrect for variables\textbf{ }$\left(
\mathbf{r},\mathbf{v}\right)  .$ Solution (\ref{19}) can be proved as follows%
\begin{equation}
\dfrac{\partial\rho\left(  \mathbf{r},\mathbf{p},t\right)  }{\partial
t}=\dfrac{\partial\rho}{\partial R_{i}}\dfrac{dR_{i}\left(  \mathbf{r}%
,\mathbf{p},t^{\prime}-t\right)  }{dt}+\dfrac{\partial\rho}{\partial P_{i}%
}\dfrac{dP_{i}\left(  \mathbf{r},\mathbf{p},t^{\prime}-t\right)  }%
{dt}.\label{21}%
\end{equation}
Since
\begin{equation}
\dfrac{dR_{i}\left(  \mathbf{r},\mathbf{p},t^{\prime}-t\right)  }{dt}%
=-\dfrac{dR_{i}\left(  \mathbf{r},\mathbf{p},t^{\prime}-t\right)  }%
{dt^{\prime}}=-\dfrac{\partial H}{\partial P_{i}}\label{22}%
\end{equation}
and
\begin{equation}
\dfrac{dP_{i}\left(  \mathbf{r},\mathbf{p},t^{\prime}-t\right)  }{dt}%
=-\dfrac{dP_{i}\left(  \mathbf{r},\mathbf{p},t^{\prime}-t\right)  }%
{dt^{\prime}}=\dfrac{\partial H}{\partial R_{i}},\label{23}%
\end{equation}
one gets%
\begin{equation}
\dfrac{\partial\rho\left(  \mathbf{r},\mathbf{p},t\right)  }{\partial
t}=-\left\{  H,\rho\right\}  _{\mathbf{R},\mathbf{P}},\label{24}%
\end{equation}
where $\left\{  H,\rho\right\}  $ is a Poisson bracket, and using invariance
of the Poisson brackets $\left\{  H,\rho\right\}  _{\mathbf{RP}}=\left\{
H,\rho\right\}  _{\mathbf{rp}}$ one arrives at Eq. (\ref{18}).

\section{\label{s4}Atom classical motion in the rotating frame}

In this section we calculate the atomic classical trajectory in the potential
consisting of the linear part and gravity gradient part%
\begin{subequations}%
\label{25}%
\begin{align}
U\left(  \mathbf{r}\right)   &  =-m\mathbf{a}\cdot\mathbf{r}+\delta U\left(
\mathbf{r}\right)  ,\label{25a}\\
\delta U\left(  \mathbf{r}\right)   &  =\dfrac{m}{4}T_{zz}\left[
r^{2}-3\left(  \mathbf{n}\cdot\mathbf{r}\right)  ^{2}\right]  , \label{25b}%
\end{align}%
\end{subequations}%
where $\mathbf{a}=\mathbf{g}-\Omega^{2}R$$\boldsymbol{\nu}$$\times\left(
\boldsymbol{\nu}\times\mathbf{n}\right)  $ is a sum of the gravity and
centrifugal accelerations, $\mathbf{\nu=\Omega}/\Omega$ is a unit vector along
the rotation frequency, $\mathbf{n=R}/R$ is a unit vector along the system's
displacement $\mathbf{R}$ from the Earth center, $T_{zz}=-2g_{z}/R,$ $g_{z}$
is the vertical component of the gravity acceleration.

We will treat the gradient term (\ref{25b}) as a perturbation. In the zero
order the atom evolves as%
\begin{subequations}%
\label{26}%
\begin{align}
\mathbf{\dot{r}} &  =\mathbf{r}\times\mathbf{\Omega}+\dfrac{\mathbf{p}}%
{m},\label{26a}\\
\mathbf{\dot{p}} &  \mathbf{=}\left(  \mathbf{p}\times\Omega\right)
+m\mathbf{a},\label{26b}%
\end{align}%
\end{subequations}%
subject to\ initial conditions $\mathbf{r}\left(  0\right)  =\mathbf{r}_{0},$
$\mathbf{p}\left(  0\right)  =\mathbf{p}_{0},$ while first-order corrections
are given by%
\begin{subequations}%
\label{27}%
\begin{align}
\delta\mathbf{p} &  =-\dfrac{m}{2}T_{zz}\mathbf{p}_{\gamma},\label{27a}\\
\delta\mathbf{r} &  =-\dfrac{1}{2}T_{zz}\mathbf{r}_{\gamma},\label{27b}%
\end{align}%
\end{subequations}%
where $\mathbf{r}_{\gamma}$ and $\mathbf{p}_{\gamma}$ evolve as%
\begin{subequations}%
\label{28}%
\begin{align}
\mathbf{\dot{r}}_{\gamma} &  =\mathbf{r}_{\gamma}\times\mathbf{\Omega
}+\mathbf{p}_{\gamma},\label{28a}\\
\mathbf{\dot{p}}_{\gamma} &  =\mathbf{p}_{\gamma}\times\mathbf{\Omega
+r}-3\mathbf{n}\left(  \mathbf{n}\cdot\mathbf{r}\right)  ,\label{28b}%
\end{align}%
\end{subequations}%
subject to initial conditions $\mathbf{r}_{\gamma}\left(  0\right)  =0,$
$\mathbf{p}_{\gamma}\left(  0\right)  =0.$

One sees that to get\ the atom trajectory it is necessary to solve equations
of the type%
\begin{equation}
\mathbf{\dot{b}}=\Omega\mathbf{b}\times\boldsymbol{\nu}+\mathbf{f}\left(
t\right)  .\label{29}%
\end{equation}
Directing temporarily $z$ axis along $\boldsymbol{\nu,}$ one finds that
variable $\xi=b_{x}-ib_{y}$ is given by%
\begin{equation}
\xi=\exp\left(  i\Omega t\right)  \xi_{0}+\int_{0}^{t}d\tau\exp\left[
i\Omega\left(  t-\tau\right)  \right]  \left[  f_{x}\left(  \tau\right)
-if_{y}\left(  \tau\right)  \right]  ,\label{30}%
\end{equation}
so that the solution of Eq. (\ref{29}) can be found as%
\begin{gather}
\mathbf{b}=\cos\Omega t\left(  b_{0x},b_{0y},0\right)  +\sin\Omega t\left(
b_{0y},-b_{0x},0\right)  \nonumber\\
+\int_{0}^{t}d\tau\left\{  \cos\Omega\left(  t-\tau\right)  \left[
f_{x}\left(  \tau\right)  ,f_{y}\left(  \tau\right)  ,0\right]  +\sin
\Omega\left(  t-\tau\right)  \left[  f_{y}\left(  \tau\right)  ,-f_{x}\left(
\tau\right)  ,0\right]  \right\}  +\left[  0,0,b_{0z}+\int_{0}^{t}d\tau
f_{z}\left(  \tau\right)  \right]  .\label{31}%
\end{gather}
Since for the chosen coordinate system $\left(  b_{0x},b_{0y},0\right)
=-$$\boldsymbol{\nu}$$\times\left(  \boldsymbol{\nu}\times\mathbf{b}\right)  $
and $\left(  b_{0y},-b_{0x},0\right)  =-$$\boldsymbol{\nu}$$\times\mathbf{b,}$
then in the vector representation the solution of Eq. (\ref{29}) is given by%
\begin{subequations}%
\label{32}%
%

\begin{align}
\mathbf{b}  &  =\mathbf{b}_{h}+\mathbf{b}_{p},\label{32a}\\
\mathbf{b}_{h}  &  =-\boldsymbol{\nu}\times\left(  \boldsymbol{\nu}%
\times\mathbf{b}_{0}\cos\Omega t+\mathbf{b}_{0}\sin\Omega t\right)
+\boldsymbol{\nu}\left(  \boldsymbol{\nu}\cdot\mathbf{b}_{0}\right)
,\label{32b}\\
\mathbf{b}_{p}  &  =\int_{0}^{t}d\tau\left\{  -\boldsymbol{\nu}\times\left[
\boldsymbol{\nu}\times\mathbf{f}\left(  \tau\right)  \cos\Omega\left(
t-\tau\right)  +\mathbf{f}\left(  \tau\right)  \sin\Omega\left(
t-\tau\right)  \right]  +\boldsymbol{\nu}\left[  \boldsymbol{\nu}%
\cdot\mathbf{f}\left(  \tau\right)  \right]  \right\}  . \label{32c}%
\end{align}%
\end{subequations}%
If in turn the driving term in Eq. (\ref{29}) can be represented as
\begin{equation}
\mathbf{f}\left(  \tau\right)  =\boldsymbol{\nu}\times\left[  \boldsymbol{\nu
}\times\mathbf{s}q_{1}\left(  \tau\right)  +\mathbf{s}q_{2}\left(
\tau\right)  \right]  +\boldsymbol{\nu}\left(  \boldsymbol{\nu}\cdot
\mathbf{s}\right)  q_{0}\left(  \tau\right)  , \label{33}%
\end{equation}
then
\begin{align}
\mathbf{b}_{p}  &  =\boldsymbol{\nu}\times\left\{  \boldsymbol{\nu}\times
\int_{0}^{t}d\tau\mathbf{s}\left[  q_{1}\left(  \tau\right)  \cos\Omega\left(
t-\tau\right)  -q_{2}\left(  \tau\right)  \sin\Omega\left(  t-\tau\right)
\right]  \right. \nonumber\\
&  +\left.  \int_{0}^{t}d\tau\mathbf{s}\left[  q_{1}\left(  \tau\right)
\sin\Omega\left(  t-\tau\right)  +q_{2}\left(  \tau\right)  \cos\Omega\left(
t-\tau\right)  \right]  \right\}  +\boldsymbol{\nu}\int_{0}^{t}d\tau\left(
\boldsymbol{\nu}\cdot\mathbf{s}\right)  q_{0}\left(  \tau\right)  . \label{34}%
\end{align}

Consecutive application of Eqs. (\ref{32})-(\ref{34}) brought us to the
following expression for the function $\mathbf{R}\left(  \mathbf{r,p}%
,t\right)  $%
\begin{subequations}%
\label{35}%
\begin{equation}
\mathbf{R}\left(  \mathbf{r},\mathbf{p},t\right)  =\mathbf{R}_{0}\left(
\mathbf{r},\mathbf{p},t\right)  +\delta\mathbf{R}\left(  \mathbf{r}%
,\mathbf{p},t\right)  ;\label{35a}%
\end{equation}%
\begin{align}
\mathbf{R}_{0}\left(  \mathbf{r},\mathbf{p},t\right)   &  =\cos x\left(
\mathbf{r}+\dfrac{\mathbf{p}}{m}t\right)  -\sin x\boldsymbol{\nu}\times\left(
\mathbf{r}+\dfrac{\mathbf{p}}{m}t\right)  +\left(  1-\cos x\right)
\boldsymbol{\nu}\left[  \boldsymbol{\nu}\cdot\left(  \mathbf{r}+\dfrac
{\mathbf{p}}{m}t\right)  \right]  \nonumber\\
&  +t^{2}\left\{  \dfrac{\mathbf{a}}{2}-\boldsymbol{\nu}\times\left[
\boldsymbol{\nu}\times\mathbf{a}f_{0,1}\left(  x\right)  +\mathbf{a}%
f_{0,2}\left(  x\right)  \right]  \right\}  ,\label{35b}\\
f_{0,1}\left(  x\right)   &  =x^{-2}\left(  \cos x+x\sin x-1\right)
-\dfrac{1}{2},\nonumber\\
f_{0,2}\left(  x\right)   &  =x^{-2}\left(  \sin x-x\cos x\right)  ;\nonumber
\end{align}%
\begin{equation}
\delta\mathbf{R}\left(  \mathbf{r},\mathbf{p},t\right)  =-\dfrac{1}{2}%
T_{zz}\mathbf{r}_{\gamma};\quad\mathbf{r}_{\gamma}=\mathbf{r}_{1}%
+\ldots\mathbf{r}_{8};\label{35c}%
\end{equation}%
\begin{align}
\mathbf{r}_{1} &  =t^{2}\left\{  \boldsymbol{\nu}\left[  \boldsymbol{\nu}%
\cdot\left(  \dfrac{\mathbf{r}}{2}+\dfrac{\mathbf{p}}{6m}t\right)  \right]
\left(  1-\cos x\right)  +\left(  \dfrac{\mathbf{r}}{2}+\dfrac{\mathbf{p}}%
{6m}t\right)  \cos x-\boldsymbol{\nu}\times\left(  \dfrac{\mathbf{r}}%
{2}+\dfrac{\mathbf{p}}{6m}t\right)  \sin x\right\}  \nonumber\\
&  +t^{4}\left\{  \mathbf{a}f_{1,1}\left(  x\right)  +\boldsymbol{\nu}\left(
\mathbf{a}\cdot\boldsymbol{\nu}\right)  \left[  \dfrac{1}{24}-f_{1,1}\left(
x\right)  \right]  -\boldsymbol{\nu}\times\mathbf{a}f_{1,2}\left(  x\right)
\right\}  ,\label{35d}\\
f_{1,1}\left(  x\right)   &  =x^{-4}\left[  1-\cos x\left(  1-2^{-1}%
x^{2}\right)  -x\sin x\left(  1-6^{-1}x^{2}\right)  \right]  ,\nonumber\\
f_{1,2}\left(  x\right)   &  =x^{-4}\left[  x\cos x\left(  1-6^{-1}%
x^{2}\right)  -\sin x\left(  1-2^{-1}x^{2}\right)  \right]  ;\nonumber
\end{align}%
\begin{align}
\mathbf{r}_{2} &  =3t^{2}\left(  \mathbf{n}\cdot\boldsymbol{\nu}\right)
\left\{  \boldsymbol{\nu}\left(  \mathbf{n}\cdot\boldsymbol{\nu}\right)
\left[  \left(  \boldsymbol{\nu}\cdot\mathbf{r}\right)  \left(  f_{21}\left(
x\right)  -\dfrac{1}{2}\right)  +t\left(  \boldsymbol{\nu}\cdot\dfrac
{\mathbf{p}}{m}\right)  \left[  f_{22}\left(  x\right)  -6^{-1}\right]
+t^{2}\boldsymbol{\nu}\cdot\mathbf{a}\left(  f_{23}\left(  x\right)
-\dfrac{1}{24}\right)  \right]  \right.  \nonumber\\
&  \left.  -\mathbf{n}\left[  \boldsymbol{\nu}\cdot\left(  \mathbf{r}%
f_{21}\left(  x\right)  +\dfrac{\mathbf{p}}{m}tf_{22}\left(  x\right)
+\mathbf{a}t^{2}f_{23}\left(  x\right)  \right)  \right]  +\boldsymbol{\nu
}\times\mathbf{n}\left[  \boldsymbol{\nu}\cdot\left(  \mathbf{r}f_{02}\left(
x\right)  +\dfrac{\mathbf{p}}{m}tf_{25}\left(  x\right)  +\mathbf{a}%
t^{2}f_{26}\left(  x\right)  \right)  \right]  \right\}  ,\label{35e}\\
f_{21}\left(  x\right)   &  =x^{-2}\left(  \cos x+x\sin x-1\right)
,\nonumber\\
f_{22}\left(  x\right)   &  =x^{-3}\left[  2\sin x-x\left(  1+\cos x\right)
\right]  ,\nonumber\\
f_{23}\left(  x\right)   &  =x^{-4}\left[  3\left(  1-\cos x\right)  -x\sin
x-2^{-1}x^{2}\right]  ,\nonumber\\
f_{25}\left(  x\right)   &  =x^{-3}\left[  2\left(  1-\cos x\right)  -x\sin
x\right]  ,\nonumber\\
f_{26}\left(  x\right)   &  =x^{-4}\left(  x\left(  2+\cos x\right)  -3\sin
x\right)  ;\nonumber
\end{align}%
\begin{align}
\mathbf{r}_{3} &  =-4^{-1}3t^{2}\left[  \left(  \boldsymbol{\nu}%
\times\mathbf{n}\right)  \cdot\left(  \boldsymbol{\nu}\times\mathbf{r}\right)
\right]  \left[  f_{30}\left(  x\right)  \boldsymbol{\nu}\left(
\mathbf{n}\cdot\boldsymbol{\nu}\right)  +f_{31}\left(  x\right)
\mathbf{n}+f_{32}\left(  x\right)  \boldsymbol{\nu}\times\mathbf{n}\right]
,\label{35f}\\
f_{30}\left(  x\right)   &  =x^{-2}\left[  4-\left(  4+x^{2}\right)  \cos
x-x\sin x\right]  ,\nonumber\\
f_{31}\left(  x\right)   &  =x^{-1}\left(  \sin x+x\cos x\right)  ,\nonumber\\
f_{32}\left(  x\right)   &  =x^{-2}\left[  x\cos x-\sin x\left(
1+x^{2}\right)  \right]  ;\nonumber
\end{align}%
\begin{align}
\mathbf{r}_{4} &  =-\dfrac{3}{4m}t^{3}\left[  \left(  \boldsymbol{\nu}%
\times\mathbf{n}\right)  \cdot\left(  \boldsymbol{\nu}\times\mathbf{p}\right)
\right]  \left[  f_{40}\left(  x\right)  \boldsymbol{\nu}\left(
\mathbf{n}\cdot\boldsymbol{\nu}\right)  +f_{41}\left(  x\right)
\mathbf{n}-\dfrac{1}{3}\sin x\boldsymbol{\nu}\times\mathbf{n}\right]
,\label{35g}\\
f_{40}\left(  x\right)   &  =x^{-3}\left[  7\sin x-x\left(  3+3^{-1}%
x^{2}\right)  \cos x-4x\right]  ,\nonumber\\
f_{41}\left(  x\right)   &  =x^{-3}\left[  \sin x-x\left(  1-3^{-1}%
x^{2}\right)  \cos x\right]  ;\nonumber
\end{align}%
\begin{align}
\mathbf{r}_{5} &  =-4^{-1}3t^{2}\left[  \left(  \boldsymbol{\nu}%
\times\mathbf{n}\right)  \cdot\mathbf{r}\right]  \left[  f_{50}\left(
x\right)  \boldsymbol{\nu}\left(  \mathbf{n}\cdot\boldsymbol{\nu}\right)
+f_{51}\left(  x\right)  \mathbf{n}-f_{02}\left(  x\right)  \boldsymbol{\nu
}\times\mathbf{n}\right]  ,\label{35h}\\
f_{50}\left(  x\right)   &  =x^{-2}\left[  x\left(  4-\cos x\right)  -\sin
x\left(  3+x^{2}\right)  \right]  ,\nonumber\\
f_{51}\left(  x\right)   &  =x^{-2}\left[  \sin x\left(  1-x^{2}\right)
-x\cos x\right]  ;\nonumber
\end{align}%
\begin{align}
\mathbf{r}_{6} &  =-\dfrac{3}{4m}t^{3}\left[  \left(  \boldsymbol{\nu}%
\times\mathbf{n}\right)  \cdot\mathbf{p}\right]  \left[  f_{60}\left(
x\right)  \boldsymbol{\nu}\left(  \mathbf{n}\cdot\boldsymbol{\nu}\right)
+3^{-1}\sin x\mathbf{n}+f_{62}\left(  x\right)  \boldsymbol{\nu}%
\times\mathbf{n}\right]  ,\label{35i}\\
f_{60}\left(  x\right)   &  =x^{-3}\left[  6\left(  1-\cos x\right)  -x\sin
x\left(  4+3^{-1}x^{2}\right)  \right]  ,\nonumber\\
f_{62}\left(  x\right)   &  =x^{-3}\left[  x\cos x\left(  1+3^{-1}%
x^{2}\right)  -\sin x\right]  ;\nonumber
\end{align}%
\begin{align}
\mathbf{r}_{7} &  =-\dfrac{3}{4}t^{4}\left[  \left(  \boldsymbol{\nu}%
\times\mathbf{n}\right)  \cdot\mathbf{a}\right]  \left[  f_{70}\left(
x\right)  \boldsymbol{\nu}\left(  \mathbf{n}\cdot\boldsymbol{\nu}\right)
+f_{71}\left(  x\right)  \mathbf{n}+f_{72}\left(  x\right)  \boldsymbol{\nu
}\times\mathbf{n}\right]  \label{35j}\\
f_{70}\left(  x\right)   &  =x^{-4}\left\{  2x\left[  4+\left(  1+6^{-1}%
x^{2}\right)  \cos x\right]  -\sin x\left(  10+x^{2}\right)  \right\}
,\nonumber\\
f_{71}\left(  x\right)   &  =x^{-4}\left[  2x\cos x\left(  1-6^{-1}%
x^{2}\right)  -\sin x\left(  2-x^{2}\right)  \right]  ,\nonumber\\
f_{72}\left(  x\right)   &  =x^{-4}\left[  x\cos x-\sin x\left(  1-3^{-1}%
x^{2}\right)  \right]  ;\nonumber
\end{align}%
\begin{align}
\mathbf{r}_{8} &  =-\dfrac{3}{4}t^{4}\left[  \left(  \boldsymbol{\nu}%
\times\mathbf{n}\right)  \cdot\left(  \boldsymbol{\nu}\times\mathbf{a}\right)
\right]  \left[  f_{8,0}\left(  x\right)  \boldsymbol{\nu}\left(
\mathbf{n}\cdot\boldsymbol{\nu}\right)  +f_{8,1}\left(  x\right)
\mathbf{n}+f_{8,2}\left(  x\right)  \boldsymbol{\nu}\times\mathbf{n}\right]
\label{35k}\\
f_{8,0}\left(  x\right)   &  =x^{-4}\left\{  8\left[  1-\left(  1+8^{-1}%
x^{2}\right)  \cos x\right]  -x\sin x\left(  1+3^{-1}x^{2}\right)
-2x^{2}\right\}  ,\nonumber\\
f_{8,1}\left(  x\right)   &  =x^{-4}\left[  4-\cos x\left(  4-x^{2}\right)
-x\sin x\left(  3-3^{-1}x^{2}\right)  \right]  ,\nonumber\\
f_{8,2}\left(  x\right)   &  =x^{-4}\left[  \sin x\left(  2-x^{2}\right)
-x\cos x\left(  2-3^{-1}x^{2}\right)  \right]  ;\nonumber
\end{align}%
\end{subequations}%
where $x=\Omega t.$

\section{\label{s5}Density matrix evolution inside Raman pulse}

Consider atoms interacting with a field of two traveling waves,%
\begin{equation}
\mathbf{E=}\dfrac{1}{2}\operatorname{Re}\left\{  \mathbf{E}_{1}e^{i\left(
\mathbf{q}_{1}\cdot\mathbf{r}-\omega_{1}t+\psi_{1}\right)  }+\mathbf{E}%
_{2}e^{i\left(  \mathbf{q}_{2}\cdot\mathbf{r}-\omega_{2}t+\psi_{2}\right)
}\right\}  ,\label{36}%
\end{equation}
where $\mathbf{E}_{j},~\omega_{j},~\mathbf{q}_{j},$ and $\psi_{j}$ are
amplitudes, frequencies,, wave vectors, and phases of waves, respectively. We
assume that the pulse of the field (\ref{36}) start and end times are $t_{0}$
and $t_{0}+\tau,$ where $\tau$ is the pulse duration, fields $\mathbf{E}_{1}$
and $\mathbf{E}_{2}$ are resonant to adjacent transitions $\left\vert
g\right\rangle \rightarrow\left\vert i\right\rangle $ and $\left\vert
e\right\rangle \rightarrow\left\vert i\right\rangle $ correspondingly, and $g$
and $e$ denote lower and upper hyperfine sublevels of the atomic ground state
manyfold, {}while $i$ corresponds to the intermediate level in the Raman
two-quantum transition. Regarding the pulse duration we assume that being
larger than inverse fields' detunings%
\begin{equation}
\tau\gg\left\vert \Delta\right\vert ^{-1},\label{37}%
\end{equation}
where
\begin{equation}
\Delta=\omega_{1}-\omega_{ig}\approx\omega_{2}-\omega_{ie},\label{38}%
\end{equation}
$\omega_{ig}$ and $\omega_{ie}$ are transition frequencies, it is also
sufficiently small in respect to all other relevant time intervals%
\begin{equation}
\tau\ll\min\left\{  T_{i},\left\vert \delta_{12}\right\vert ^{-1},\left(
k\Delta p/m\right)  ^{-1},\left(  \hbar k^{2}/m\right)  ^{-1},\left\vert
\mathbf{k}\cdot\mathbf{g}\right\vert ^{-1/2},\left\vert \Omega\right\vert
^{-1}\right\}  ,\label{39}%
\end{equation}
where $i\not =1,$
\begin{equation}
\mathbf{k}=\mathbf{q}_{1}-\mathbf{q}_{2}\label{40}%
\end{equation}
is the effective wave vector,%
\begin{equation}
\delta_{12}=\omega_{1}-\omega_{2}-\omega_{eg}\label{41}%
\end{equation}
is the Raman detuning, $\Delta p$ is the width of the atoms' distribution over momenta.

Evolution of atomic levels amplitudes under assumption (\ref{37}) and in the
absence of acceleration and rotation has been considered in review \cite{c16}.
From Eqs. (51) and (52) in \cite{c16}, using conditions (\ref{39}) one finds
that in the Shr\"{o}dinger representation amplitudes of levels after the pulse
action are given by%
\begin{subequations}%
\label{42}%
\begin{align}
c\left(  e,\mathbf{p}+\hbar\mathbf{k},t_{0}+\tau\right)   &  =\exp\left\{
-i\tau\left(  \Omega_{e}^{AC}+\Omega_{g}^{AC}\right)  /2\right\}  \left\{
\cos\left(  \dfrac{\theta}{2}\right)  c\left(  e,\mathbf{p}+\hbar
\mathbf{k},t_{0}\right)  \right.  \nonumber\\
&  \left.  -i\exp\left\{  -i\left[  \phi+\delta_{12}t_{0}\right]  \right\}
\sin\left(  \dfrac{\theta}{2}\right)  c\left(  g,\mathbf{p},t_{0}\right)
\right\}  ,\label{42a}\\
c\left(  g,\mathbf{p},t_{0}+\tau\right)   &  =\exp\left\{  -i\tau\left(
\Omega_{e}^{AC}+\Omega_{g}^{AC}\right)  /2\right\}  \left\{  -i\exp\left\{
i\left[  \phi+\delta_{12}t_{0}\right]  \right\}  \sin\left(  \dfrac{\theta}%
{2}\right)  c\left(  e,\mathbf{p}+\hbar\mathbf{k},t_{0}\right)  \right.
\nonumber\\
&  \left.  +\cos\left(  \dfrac{\theta}{2}\right)  c\left(  g,\mathbf{p}%
,t_{0}\right)  \right\}  ,\label{42b}%
\end{align}%
\end{subequations}%
where $\Omega_{e}^{AC}=\left\vert \Omega_{e}\right\vert ^{2}/4\Delta$ and
$\Omega_{g}^{AC}=\left\vert \Omega_{g}\right\vert ^{2}/4\Delta$ are ac-Stark
shifts of levels $\left\vert e\right\rangle $ and $\left\vert g\right\rangle
,$ $\Omega_{e}=-\left\langle i\left\vert \mathbf{d}\cdot\mathbf{E}%
_{2}\right\vert e\right\rangle /\hbar,~\Omega_{g}=-\left\langle i\left\vert
\mathbf{d}\cdot\mathbf{E}_{1}\right\vert g\right\rangle /\hbar,$
\begin{align}
\phi &  =\psi_{2}-\psi_{1},\label{421}\\
\theta &  =\Omega_{eff}\tau\label{43}%
\end{align}
is the pulse area, $\Omega_{eff}=\left\vert \Omega_{e}\Omega_{g}%
/2\Delta\right\vert $ is the effective Rabi frequency associated with the
Raman transition $\left\vert e\right\rangle \rightarrow\left\vert
g\right\rangle .$ In the derivation of Eqs. (\ref{42}) we assumed that Raman
detuning is much smaller than the effective Rabi frequency, $\left\vert
\delta_{12}\right\vert \ll\Omega_{eff}.$ For the optimal pulse areas
$\theta\sim1,$ this condition follows from the inequality (\ref{39}).

For the density matrix \ in the Wigner representation (\ref{12}), which can
\ be expressed through levels' amplitudes as%
\begin{equation}
\rho_{\alpha\beta}\left(  \mathbf{r},\mathbf{p},t\right)  =\int\dfrac
{d\mathbf{q}}{\left(  2\pi\right)  ^{3}}c\left(  \alpha,\mathbf{p}+\dfrac
{1}{2}\hbar\mathbf{q},t\right)  c^{\ast}\left(  \beta,\mathbf{p}-\dfrac{1}%
{2}\hbar\mathbf{q},t\right)  e^{i\mathbf{q}\cdot\mathbf{r}},\label{44}%
\end{equation}
one finds%
\begin{subequations}%
\label{45}%
\begin{align}
\rho_{ee}\left(  \mathbf{r},\mathbf{p},t_{0}+\tau\right)   &  =\cos^{2}\left(
\dfrac{\theta}{2}\right)  \rho_{ee}\left(  \mathbf{r},\mathbf{p},t_{0}\right)
+\sin^{2}\left(  \dfrac{\theta}{2}\right)  \rho_{gg}\left(  \mathbf{r}%
,\mathbf{p}-\hbar\mathbf{k},t_{0}\right)  \nonumber\\
&  +\operatorname{Re}\left\{  i\sin\theta\exp\left[  -i\left(  \mathbf{k}%
\cdot\mathbf{r}-\delta_{12}t_{0}-\phi\right)  \right]  \rho_{eg}\left(
\mathbf{r},\mathbf{p}-\dfrac{\hbar\mathbf{k}}{2},t_{0}\right)  \right\}
;\label{45a}\\
\rho_{gg}\left(  \mathbf{r},\mathbf{p},t_{0}+\tau\right)   &  =\sin^{2}\left(
\dfrac{\theta}{2}\right)  \rho_{ee}\left(  \mathbf{r},\mathbf{p}%
+\hbar\mathbf{k},t_{0}\right)  +\cos^{2}\left(  \dfrac{\theta}{2}\right)
\rho_{gg}\left(  \mathbf{r},\mathbf{p},t_{0}\right)  \nonumber\\
&  -\operatorname{Re}\left\{  i\sin\theta\exp\left[  -i\left(  \mathbf{k}%
\cdot\mathbf{r}-\delta_{12}t_{0}-\phi\right)  \right]  \rho_{eg}\left(
\mathbf{r},\mathbf{p}+\dfrac{\hbar\mathbf{k}}{2},t_{0}\right)  \right\}
;\label{45b}\\
\rho_{eg}\left(  \mathbf{r},\mathbf{p},t_{0}+\tau\right)   &  =\dfrac{1}%
{2}i\sin\theta\exp\left[  i\left(  \mathbf{k}\cdot\mathbf{r}-\delta_{12}%
t_{0}-\phi\right)  \right]  \left[  \rho_{ee}\left(  \mathbf{r},\mathbf{p}%
+\dfrac{\hbar\mathbf{k}}{2},t_{0}\right)  -\rho_{gg}\left(  \mathbf{r}%
,\mathbf{p}-\dfrac{\hbar\mathbf{k}}{2},t_{0}\right)  \right]  \nonumber\\
&  +\cos^{2}\left(  \dfrac{\theta}{2}\right)  \rho_{eg}\left(  \mathbf{r}%
,\mathbf{p},t_{0}\right)  +\sin^{2}\left(  \dfrac{\theta}{2}\right)
\exp\left[  2i\left(  \mathbf{k}\cdot\mathbf{r}-\delta_{12}t_{0}-\phi\right)
\right]  \rho_{ge}\left(  \mathbf{r},\mathbf{p},t_{0}\right)  .\label{45c}%
\end{align}%
\end{subequations}%
In the specific cases of $\pi$ and $\pi/2$ pulses Eqs. (\ref{45}) reduce to%
\begin{subequations}%
\label{46}%
\begin{align}
\rho_{ee}\left(  \mathbf{r},\mathbf{p},t_{0}+\tau\right)   &  =\rho
_{gg}\left(  \mathbf{r},\mathbf{p}-\hbar\mathbf{k},t_{0}\right)
;\label{46a}\\
\rho_{gg}\left(  \mathbf{r},\mathbf{p},t_{0}+\tau\right)   &  =\rho
_{ee}\left(  \mathbf{r},\mathbf{p}+\hbar\mathbf{k},t_{0}\right)
;\label{46b}\\
\rho_{eg}\left(  \mathbf{r},\mathbf{p},t_{0}+\tau\right)   &  =\exp\left[
2i\left(  \mathbf{k}\cdot\mathbf{r}-\delta_{12}t_{0}-\phi\right)  \right]
\rho_{ge}\left(  \mathbf{r},\mathbf{p},t_{0}\right)  ,\label{46c}%
\end{align}%
\end{subequations}%
for $\theta=\pi,$ and%
\begin{subequations}%
\label{47}
\begin{align}
\rho_{ee}\left(  \mathbf{r},\mathbf{p},t_{0}+\tau\right)   &  =\dfrac{1}%
{2}\left[  \rho_{ee}\left(  \mathbf{r},\mathbf{p},t_{0}\right)  +\rho
_{gg}\left(  \mathbf{r},\mathbf{p}-\hbar\mathbf{k},t_{0}\right)  \right]
\nonumber\\
&  +\operatorname{Re}\left\{  i\exp\left[  -i\left(  \mathbf{k}\cdot
\mathbf{r}-\delta_{12}t_{0}-\phi\right)  \right]  \rho_{eg}\left(
\mathbf{r},\mathbf{p}-\dfrac{\hbar\mathbf{k}}{2},t_{0}\right)  \right\}
;\label{47a}\\
\rho_{gg}\left(  \mathbf{r},\mathbf{p},t_{0}+\tau\right)   &  =\dfrac{1}%
{2}\left[  \rho_{ee}\left(  \mathbf{r},\mathbf{p}+\hbar\mathbf{k}%
,t_{0}\right)  +\rho_{gg}\left(  \mathbf{r},\mathbf{p},t_{0}\right)  \right]
\nonumber\\
&  -\operatorname{Re}\left\{  i\exp\left[  -i\left(  \mathbf{k}\cdot
\mathbf{r}-\delta_{12}t_{0}-\phi\right)  \right]  \rho_{eg}\left(
\mathbf{r},\mathbf{p}+\dfrac{\hbar\mathbf{k}}{2},t_{0}\right)  \right\}
;\label{47b}\\
\rho_{eg}\left(  \mathbf{r},\mathbf{p},t_{0}+\tau\right)   &  =\dfrac{i}%
{2}\exp\left[  i\left(  \mathbf{k}\cdot\mathbf{r}-\delta_{12}t_{0}%
-\phi\right)  \right]  \left[  \rho_{ee}\left(  \mathbf{r},\mathbf{p}%
+\dfrac{\hbar\mathbf{k}}{2},t_{0}\right)  -\rho_{gg}\left(  \mathbf{r}%
,\mathbf{p}-\dfrac{\hbar\mathbf{k}}{2},t_{0}\right)  \right]  \nonumber\\
&  +\dfrac{1}{2}\left\{  \rho_{eg}\left(  \mathbf{r},\mathbf{p},t_{0}\right)
+\exp\left[  2i\left(  \mathbf{k}\cdot\mathbf{r}-\delta_{12}t_{0}-\phi\right)
\right]  \rho_{ge}\left(  \mathbf{r},\mathbf{p},t_{0}\right)  \right\}
\label{47c}%
\end{align}%
\end{subequations}%
for $\theta=\pi/2.$

\section{\label{s6}Three pulse atom interferometer}

Consider first a $\pi/2-\pi-\pi/2$ sequence of pulses applied at moments
$T_{1}=0,~T_{2}=T_{3}/2,~T_{3}$. If initially all atoms are in the lower
state, where their distribution in the Wigner representation is $f\left(
\mathbf{r},\mathbf{p}\right)  ,$%
\begin{equation}
\rho\left(  \mathbf{r},\mathbf{p,0}\right)  =\left(
\begin{array}
[c]{cc}%
0 & 0\\
0 & f\left(  \mathbf{r},\mathbf{p}\right)
\end{array}
\right)  ,\label{48}%
\end{equation}
then applying consequently Eqs. (\ref{19}), (\ref{46}), (\ref{47}) one finds
that after the third pulse action, the upper state distribution is given by%
\begin{align}
\rho_{ee}\left(  \mathbf{r},\mathbf{p},T_{3}+\tau\right)   &  =\dfrac{1}%
{4}\left\{  f\left[  \mathbf{R}\left(  \mathbf{R}\left(  \mathbf{r}%
,\mathbf{p},T_{2}-T_{3}\right)  ,\mathbf{P}\left(  \mathbf{r},\mathbf{p}%
,T_{2}-T_{3}\right)  -\hbar\mathbf{k},-T_{2}\right)  ,\right.  \right.
\nonumber\\
&  \left.  \mathbf{P}\left(  \mathbf{R}\left(  \mathbf{r},\mathbf{p}%
,T_{2}-T_{3}\right)  ,\mathbf{P}\left(  \mathbf{r},\mathbf{p},T_{2}%
-T_{3}\right)  -\hbar\mathbf{k},-T_{2}\right)  \right]  \nonumber\\
&  +f\left[  \mathbf{R}\left(  \mathbf{R}\left(  \mathbf{r},\mathbf{p}%
-\hbar\mathbf{k},T_{2}-T_{3}\right)  ,\mathbf{P}\left(  \mathbf{r}%
,\mathbf{p}-\hbar\mathbf{k},T_{2}-T_{3}\right)  +\hbar\mathbf{k}%
,-T_{2}\right)  ,\right.  \nonumber\\
&  \left.  \left.  \mathbf{P}\left(  \mathbf{R}\left(  \mathbf{r}%
,\mathbf{p}-\hbar\mathbf{k},T_{2}-T_{3}\right)  ,\mathbf{P}\left(
\mathbf{r},\mathbf{p}-\hbar\mathbf{k},T_{2}-T_{3}\right)  +\hbar
\mathbf{k},-T_{2}\right)  -\hbar\mathbf{k}\right]  \right\}  \nonumber\\
&  -\dfrac{1}{2}\cos\left[  \mathbf{k}\cdot\mathbf{r}-2\mathbf{k}%
\cdot\mathbf{R}\left(  \mathbf{r},\mathbf{p}-\dfrac{\hbar\mathbf{k}}{2}%
,T_{2}-T_{3}\right)  +\mathbf{k}\cdot\mathbf{R}\left(  \mathbf{r}%
,\mathbf{p}-\dfrac{\hbar\mathbf{k}}{2},-T_{3}\right)  \right.  \nonumber\\
&  \left.  -\delta_{12}\left(  T_{3}-2T_{2}\right)  -\phi_{3}+2\phi_{2}%
-\phi_{1}\right]  f\left[  \mathbf{R}\left(  \mathbf{r},\mathbf{p}%
-\dfrac{\hbar\mathbf{k}}{2},-T_{3}\right)  ,\mathbf{P}\left(  \mathbf{r}%
,\mathbf{p}-\dfrac{\hbar\mathbf{k}}{2},-T_{3}\right)  -\dfrac{\hbar\mathbf{k}%
}{2}\right]  ;\label{49}%
\end{align}
This expression can be used to calculate any response associated with atoms on
the upper level. It can be used to study the spatial shift and deformation of
atomic clouds. In this paper we use it to get the total probability of atoms
excitation%
\begin{equation}
w\equiv\int d\mathbf{r}d\mathbf{p}\rho_{ee}\left(  \mathbf{r},\mathbf{p}%
,T_{3}+\tau\right)  .\label{50}%
\end{equation}
Choosing for each term in Eq. (\ref{49}) arguments of the initial distribution
$f$ as integration variables and taking into account that phase space stays
invariant under both the free atom motion and recoil kicks of the momentum,
one arrives at the following probability of excitation:%
\begin{subequations}%
\label{51}%
\begin{align}
w &  =\dfrac{1}{2}\left[  1-\int d\mathbf{r}d\mathbf{p}\cos\left(  \phi
_{m}-\phi_{R}-\phi_{3}-\phi_{1}+2\phi_{2}\right)  f\left(  \mathbf{r}%
,\mathbf{p}\right)  \right]  ,\label{51a}\\
\phi_{m} &  =\mathbf{k}\cdot\left[  \mathbf{R}\left(  \mathbf{r}%
,\mathbf{p}+\dfrac{\hbar\mathbf{k}}{2},T_{3}\right)  -2\mathbf{R}\left(
\mathbf{r},\mathbf{p}+\dfrac{\hbar\mathbf{k}}{2},T_{2}\right)  +\mathbf{r}%
\right]  ,\label{51b}\\
\phi_{R} &  =\delta_{12}\left(  T_{3}-2T_{2}\right)  ,\label{51c}%
\end{align}%
\end{subequations}%
where $\phi_{j}$ is a phase (\ref{421}) associated with the pulse $j,$ and
$\phi_{m}$ is the phase caused by atomic motion and recoil effect, which
reduces to the Doppler and recoil phase shifts in the absence of rotation and
acceleration, and $\phi_{R}$ is Ramsey phase \cite{c17}.

Time independent and linear in $T_{j}$ parts of the $\phi_{m}$ cancel owing to
the coincidence of wave vectors of pulses (phase-matching conditions) and
proper relations between pulses' separations. Other terms can be mapped out as
those depending on $\mathbf{r}$ and $\mathbf{r}$ independent. We assume that
the atomic cloud size is sufficiently small to neglect $\mathbf{r}$-dependent
terms. We accept the same assumption for nonlinear in $T_{j}$ parts of the
$\phi_{m}$ regarding the momentum distribution, except that we allow atoms to
be launched into some initial momentum $\mathbf{p}$. Thereupon one can drop
integration in Eq. (\ref{51a}) over phase space, and $\phi_{m}$ becomes the
phase of the interferometer.

All phase terms up to the order $T_{3}^{4}$ are presented in Table \ref{t1}.
They can be obtained by multiplying by $\mathbf{k}$ corresponding vectors in
Eqs. (\ref{35}), expanding functions $f_{ij}\left(  x\right)  $ and replacing
factor $t^{n}$ $\left(  n>1\right)  $ as%
\begin{equation}
t^{n}\rightarrow T_{3}^{n}\left(  1-2^{1-n}\right)  .\label{52}%
\end{equation}
We would like to mention that the fourth term in this table is 1.5 times
smaller than the corresponding part of the phase calculated in \cite{c72},
which has to be corrected.

As an example, and to verify our calculations we present in columns 2 and 3
phase terms for the case considered in article \cite{c8}, when atoms are
launched from the coordinate system origin and gravity acceleration, launch
velocity, and wave vectors are all vertical,%
\begin{align}
\mathbf{g} &  \mathbf{=}g_{z}\mathbf{n,}\nonumber\\
\mathbf{k} &  \mathbf{=}k\mathbf{n,}\nonumber\\
\mathbf{p} &  =mv_{z}\mathbf{n.}\label{521}%
\end{align}
The choice of the coordinate axes are also the same, $z,$ vertical; $y,$
south-north; $x,$ west-east directions.

\begin{table}[ptb]
\caption{Three-pulse interferometer phase shifts. Column 1, for an arbitrary
acceleration $\mathbf{a,}$ rotation frequency $\boldsymbol{\Omega,}$ and wave
vector $\mathbf{k;}$ column 2, for the vertical gravity acceleration $\left(
\mathbf{g=}g_{z}\mathbf{n}\right)  \mathbf{,}$ launch velocity $\left(
\mathbf{p}=mv_{z}\mathbf{n}\right)  $ and wave vector $\left(  \mathbf{k}%
=k\mathbf{n}\right)  $. Those dependencies which appear in the different cells
of column 2 are underlined and summed up in the column 3.}%
\begin{tabular}
[c]{l}\hline\hline
$%
\begin{array}
[c]{ccccc}%
\smallskip\dfrac{1}{4}\mathbf{k}\cdot\mathbf{a}T_{3}^{2} & \hspace{0.75in} &
\begin{array}
[c]{c}%
\smallskip\dfrac{1}{4}kg_{z}T_{3}^{2}\\
\dfrac{1}{4}k\Omega_{y}^{2}RT_{3}^{2}%
\end{array}
& \hspace{0.75in} & \\
\smallskip\dfrac{1}{2m}\left(  \boldsymbol{\Omega}\times\mathbf{k}\right)
\cdot\mathbf{p}T_{3}^{2} & \hspace{0.75in} & 0 & \hspace{0.5in} & \\
\smallskip\dfrac{3\Omega^{2}}{8m}\left[  \left(  \mathbf{k}\boldsymbol{\nu
}\right)  \left(  \mathbf{p}\boldsymbol{\nu}\right)  -\left(  \mathbf{kp}%
\right)  \right]  T_{3}^{3} & \hspace{0.5in} & -\dfrac{3}{8}kv_{z}\Omega
_{y}^{2}T_{3}^{3} & \hspace{0.5in} & \\
\smallskip\dfrac{1}{4}\boldsymbol{\Omega}\cdot\left(  \mathbf{k}%
\times\mathbf{a}\right)  T_{3}^{3} & \hspace{0.5in} & 0 & \hspace{0.5in} & \\
\smallskip\dfrac{3\hbar}{16m}\left[  \left(  \mathbf{k}\cdot\boldsymbol{\nu
}\right)  ^{2}-k^{2}\right]  \Omega^{2}T_{3}^{3} & \hspace{0.5in} &
-\dfrac{3\hbar k^{2}}{16m}\Omega_{y}^{2}T_{3}^{3} & \hspace{0.5in} & \\
\smallskip-\dfrac{7}{64}\left[  \left(  \mathbf{k}\cdot\mathbf{a}\right)
-\left(  \mathbf{k}\cdot\boldsymbol{\nu}\right)  \left(  \mathbf{a}%
\cdot\boldsymbol{\nu}\right)  \right]  \Omega^{2}T_{3}^{4} & \hspace{0.5in} &
\begin{array}
[c]{c}%
\smallskip-\dfrac{7}{64}kg_{z}\Omega_{y}^{2}T_{3}^{4}\\
-\dfrac{7}{64}k\Omega^{2}\Omega_{y}^{2}RT_{3}^{4}%
\end{array}
& \hspace{0.5in} & \\
\smallskip-\dfrac{7}{48m}\left(  \boldsymbol{\nu}\times\mathbf{k}\right)
\cdot\mathbf{p}\Omega^{3}T_{3}^{4} & \hspace{0.5in} & 0 & \hspace{0.5in} & \\
\smallskip-\dfrac{1}{16m}\mathbf{k}\cdot\mathbf{p}T_{3}^{3}T_{zz} &
\hspace{0.5in} & \underrightarrow{-\dfrac{1}{16}kv_{z}T_{3}^{3}T_{zz}} &
\hspace{0.5in} & \dfrac{1}{8}kv_{z}T_{3}^{3}T_{zz}\\
\smallskip-\dfrac{\hbar k^{2}}{32m}T_{3}^{3}T_{zz} & \hspace{0.5in} &
\underleftrightarrow{-\dfrac{\hbar k^{2}}{32m}T_{3}^{3}T_{zz}} &
\hspace{0.5in} & \dfrac{\hbar k^{2}}{16m}T_{3}^{3}T_{zz}\\
\smallskip-\dfrac{7}{384}\mathbf{k}\cdot\mathbf{a}T_{3}^{4}T_{zz} &
\hspace{0.5in} &
\begin{array}
[c]{c}%
\smallskip\underline{-\dfrac{7}{384}kg_{z}T_{3}^{4}T_{zz}}\\
\underleftarrow{-\dfrac{7}{384}k\Omega_{y}^{2}RT_{3}^{4}T_{zz}}%
\end{array}
& \hspace{0.5in} &
\begin{array}
[c]{c}%
\smallskip\dfrac{7}{192}kg_{z}T_{3}^{4}T_{zz}\\
\dfrac{7}{192}k\Omega_{y}^{2}RT_{3}^{4}T_{zz}%
\end{array}
\\
\smallskip\dfrac{3}{16m}\left(  \mathbf{n}\cdot\boldsymbol{\nu}\right)
\left(  \mathbf{k}\cdot\mathbf{n}\right)  \left(  \boldsymbol{\nu}%
\cdot\mathbf{p}\right)  T_{3}^{3}T_{zz} & \hspace{0.5in} & \underrightarrow
{\dfrac{3}{16}\nu_{z}^{2}kv_{z}T_{3}^{3}T_{zz}} & \hspace{0.5in} & \\
\smallskip\dfrac{3\hbar}{32m}\left(  \mathbf{n}\cdot\boldsymbol{\nu}\right)
\left(  \mathbf{k}\cdot\mathbf{n}\right)  \left(  \boldsymbol{\nu}%
\cdot\mathbf{k}\right)  T_{3}^{3}T_{zz} & \hspace{0.5in} &
\underleftrightarrow{\dfrac{3\hbar k^{2}}{32m}\nu_{z}^{2}T_{3}^{3}T_{zz}} &
\hspace{0.5in} & \\
\smallskip\dfrac{7}{128}\left(  \mathbf{n}\cdot\boldsymbol{\nu}\right)
\left(  \mathbf{k}\cdot\mathbf{n}\right)  \left(  \boldsymbol{\nu}%
\cdot\mathbf{a}\right)  T_{3}^{4}T_{zz} & \hspace{0.5in} & \underline
{\dfrac{7}{128}\nu_{z}^{2}kg_{z}T_{3}^{4}T_{zz}} & \hspace{0.5in} & \\
\smallskip-\dfrac{7}{64m}\left(  \mathbf{n}\cdot\boldsymbol{\nu}\right)
\left[  \boldsymbol{\nu}\cdot\left(  \mathbf{n}\times\mathbf{k}\right)
\right]  \left(  \boldsymbol{\nu}\cdot\mathbf{p}\right)  \Omega T_{3}%
^{4}T_{zz} & \hspace{0.5in} & 0 & \hspace{0.5in} & \\
\smallskip-\dfrac{7\hbar}{128m}\left(  \mathbf{n}\cdot\boldsymbol{\nu}\right)
\left[  \boldsymbol{\nu}\cdot\left(  \mathbf{n}\times\mathbf{k}\right)
\right]  \left(  \boldsymbol{\nu}\cdot\mathbf{k}\right)  \Omega T_{3}%
^{4}T_{zz} & \hspace{0.5in} & 0 & \hspace{0.5in} & \\
\smallskip\dfrac{3}{16m}\left[  \left(  \boldsymbol{\nu}\times\mathbf{n}%
\right)  \cdot\left(  \boldsymbol{\nu}\times\mathbf{p}\right)  \right]
\left(  \mathbf{k}\cdot\mathbf{n}\right)  T_{3}^{3}T_{zz} & \hspace{0.5in} &
\underrightarrow{\dfrac{3}{16}\nu_{y}^{2}kv_{z}T_{3}^{3}T_{zz}} &
\hspace{0.5in} & \\
\smallskip\dfrac{3\hbar}{32m}\left[  \left(  \boldsymbol{\nu}\times
\mathbf{n}\right)  \cdot\left(  \boldsymbol{\nu}\times\mathbf{k}\right)
\right]  \left(  \mathbf{k}\cdot\mathbf{n}\right)  T_{3}^{3}T_{zz} &
\hspace{0.5in} & \underleftrightarrow{\dfrac{3\hbar k^{2}}{32m}\nu_{y}%
^{2}T_{3}^{3}T_{zz}} & \hspace{0.5in} & \\
\smallskip-\dfrac{7}{64m}\left[  \left(  \boldsymbol{\nu}\times\mathbf{n}%
\right)  \cdot\left(  \boldsymbol{\nu}\times\mathbf{p}\right)  \right]
\left[  \boldsymbol{\nu}\cdot\left(  \mathbf{n}\times\mathbf{k}\right)
\right]  \Omega T_{3}^{4}T_{zz} & \hspace{0.5in} & 0 & \hspace{0.5in} & \\
\smallskip-\dfrac{7\hbar}{128m}\left[  \left(  \boldsymbol{\nu}\times
\mathbf{n}\right)  \cdot\left(  \boldsymbol{\nu}\times\mathbf{k}\right)
\right]  \left[  \boldsymbol{\nu}\cdot\left(  \mathbf{n}\times\mathbf{k}%
\right)  \right]  \Omega T_{3}^{4}T_{zz} & \hspace{0.5in} & 0 & \hspace{0.5in}
& \\
\smallskip\dfrac{7}{64m}\left[  \left(  \boldsymbol{\nu}\times\mathbf{n}%
\right)  \cdot\mathbf{p}\right]  \left(  \mathbf{k}\cdot\mathbf{n}\right)
\Omega T_{3}^{4}T_{zz} & \hspace{0.5in} & 0 & \hspace{0.5in} & \\
\smallskip\dfrac{7\hbar}{128m}\left[  \left(  \boldsymbol{\nu}\times
\mathbf{n}\right)  \cdot\mathbf{k}\right]  \left(  \mathbf{k}\cdot
\mathbf{n}\right)  \Omega T_{3}^{4}T_{zz} & \hspace{0.5in} & 0 &
\hspace{0.5in} & \\
\smallskip\dfrac{7}{128}\left[  \left(  \boldsymbol{\nu}\times\mathbf{n}%
\right)  \cdot\left(  \boldsymbol{\nu}\times\mathbf{a}\right)  \right]
\left(  \mathbf{k}\cdot\mathbf{n}\right)  T_{3}^{4}T_{zz} & \hspace{0.5in} &
\begin{array}
[c]{c}%
\smallskip\underline{\dfrac{7}{128}k\nu_{y}^{2}g_{z}T_{3}^{4}T_{zz}}\\
\underleftarrow{\dfrac{7}{128}k\nu_{y}^{2}\Omega^{2}RT_{3}^{4}T_{zz}}%
\end{array}
& \hspace{0.5in} &
\end{array}
$\\\hline\hline
\end{tabular}
\label{t1}%
\end{table}

{}If Raman pulses are comprised from noncounterpropagating fields then
effective wave vectors still have to be parallel to one another, and for any
ratio $T_{2}/T_{3}$ one can find appropriate magnitudes of absolute values of
effective wave vectors and for $T_{2}\not =T_{3}/2$ one gets nonzero Ramsey
phase (\ref{51c}).

For other values of pulses' areas, using Eqs. (\ref{45}) one arrives at the
expression for the excitation probability consisting of 10 terms. Among them
five terms associated with transferring coherences between different pairs of
pulses correspond to Ramsey fringes \cite{c17} and play no roll on Doppler
broadened transitions \cite{c18}. Another four terms are originated from the
population transfer between fields \cite{c19}; they comprise the background of
the signal. Only one term is responsible for the atom interference. Piecing
together all these terms one arrives at the following expression for the
excitation probability:%
\begin{equation}
w=\dfrac{1}{2}\left[  1-\cos\theta_{1}\cos\theta_{2}\cos\theta_{3}-\sin
\theta_{3}\sin^{2}\dfrac{\theta_{2}}{2}\sin\theta_{1}\cos\left(  \phi_{m}%
-\phi_{R}-\phi_{3}-\phi_{1}+2\phi_{2}\right)  \right]  ,\label{53}%
\end{equation}
where $\theta_{j}$ is an area of the pulse $j.$

\section{\label{s7}Four-pulse atomic sensors}

Consider now the interaction of atoms with the sequence of four pulses
\begin{equation}
\left(  \pi/2,~T_{1}=0,~\mathbf{k}_{1}\right)  -\left(  \pi,~T_{2}%
,~\mathbf{k}_{2}\right)  -\left(  \pi,~T_{3},~\mathbf{k}_{3}\right)  -\left(
\pi/2,~T_{4},~\mathbf{k}_{4}\right)  .\label{54}%
\end{equation}
Using Eqs. (\ref{19}), (\ref{46}), and (\ref{47}) one finds the upper state
distribution
\begin{align}
\rho_{ee}\left(  \mathbf{r},\mathbf{p},T_{4}+\tau\right)   &  =\dfrac{1}%
{4}\left\{  f\left[  \mathbf{R}\left(  \mathbf{R}\left(  \mathbf{R}\left(
\mathbf{r},\mathbf{p},T_{3}-T_{4}\right)  ,\mathbf{P}\left(  \mathbf{r}%
,\mathbf{p},T_{3}-T_{4}\right)  -\hbar\mathbf{k}_{3},T_{2}-T_{3}\right)
,\right.  \right.  \right.  \nonumber\\
&  \left.  \mathbf{P}\left(  \mathbf{R}\left(  \mathbf{r},\mathbf{p}%
,T_{3}-T_{4}\right)  ,\mathbf{P}\left(  \mathbf{r},\mathbf{p},T_{3}%
-T_{4}\right)  -\hbar\mathbf{k}_{3},T_{2}-T_{3}\right)  +\hbar\mathbf{k}%
_{1},-T_{2}\right)  ,\nonumber\\
&  \mathbf{P}\left(  \mathbf{R}\left(  \mathbf{R}\left(  \mathbf{r}%
,\mathbf{p},T_{3}-T_{4}\right)  ,\mathbf{P}\left(  \mathbf{r},\mathbf{p}%
,T_{3}-T_{4}\right)  -\hbar\mathbf{k}_{3},T_{2}-T_{3}\right)  ,\right.
\nonumber\\
&  \left.  \left.  \mathbf{P}\left(  \mathbf{R}\left(  \mathbf{r}%
,\mathbf{p},T_{3}-T_{4}\right)  ,\mathbf{P}\left(  \mathbf{r},\mathbf{p}%
,T_{3}-T_{4}\right)  -\hbar\mathbf{k}_{3},T_{2}-T_{3}\right)  +\hbar
\mathbf{k}_{1},-T_{2}\right)  -\hbar\mathbf{k}_{1}\right]  \nonumber\\
&  +f\left[  \mathbf{R}\left(  \mathbf{R}\left(  \mathbf{R}\left(
\mathbf{r},\mathbf{p}-\hbar\mathbf{k}_{4},T_{3}-T_{4}\right)  ,\mathbf{P}%
\left(  \mathbf{r},\mathbf{p}-\hbar\mathbf{k}_{4},T_{3}-T_{4}\right)
+\hbar\mathbf{k}_{3},T_{2}-T_{3}\right)  ,\right.  \right.  \nonumber\\
&  \left.  \mathbf{P}\left(  \mathbf{R}\left(  \mathbf{r},\mathbf{p}%
-\hbar\mathbf{k}_{4},T_{3}-T_{4}\right)  ,\mathbf{P}\left(  \mathbf{r}%
,\mathbf{p}-\hbar\mathbf{k}_{4},T_{3}-T_{4}\right)  +\hbar\mathbf{k}_{3}%
,T_{2}-T_{3}\right)  -\hbar\mathbf{k}_{2},-T_{2}\right)  ,\nonumber\\
&  \mathbf{P}\left(  \mathbf{R}\left(  \mathbf{R}\left(  \mathbf{r}%
,\mathbf{p}-\hbar\mathbf{k}_{4},T_{3}-T_{4}\right)  ,\mathbf{P}\left(
\mathbf{r},\mathbf{p}-\hbar\mathbf{k}_{4},T_{3}-T_{4}\right)  +\hbar
\mathbf{k}_{3},T_{2}-T_{3}\right)  ,\right.  \nonumber\\
&  \left.  \left.  \left.  \mathbf{P}\left(  \mathbf{R}\left(  \mathbf{r}%
,\mathbf{p}-\hbar\mathbf{k}_{4},T_{3}-T_{4}\right)  ,\mathbf{P}\left(
\mathbf{r},\mathbf{p}-\hbar\mathbf{k}_{4},T_{3}-T_{4}\right)  +\hbar
\mathbf{k}_{3},T_{2}-T_{3}\right)  -\hbar\mathbf{k}_{2},-T_{2}\right)
\right]  \right\}  \nonumber\\
&  +\dfrac{1}{2}\cos\left[  \mathbf{k}_{4}\cdot\mathbf{r}-2\mathbf{k}_{3}%
\cdot\mathbf{R}\left(  \mathbf{r},\mathbf{p},T_{3}-T_{4}\right)
+2\mathbf{k}_{2}\cdot\mathbf{R}\left(  \mathbf{r},\mathbf{p},T_{2}%
-T_{4}\right)  -\mathbf{k}_{1}\cdot\mathbf{R}\left(  \mathbf{r},\mathbf{p}%
,-T_{4}\right)  \right.  \nonumber\\
&  \left.  -\delta_{12}\left(  T_{4}-2T_{3}+2T_{2}\right)  -\left(  \phi
_{4}-2\phi_{3}+2\phi_{2}-\phi_{1}\right)  \right]  f\left[  \mathbf{R}\left(
\mathbf{r},\mathbf{p},-T_{4}\right)  ,\mathbf{P}\left(  \mathbf{r}%
,\mathbf{p},-T_{4}\right)  -\dfrac{\hbar\mathbf{k}_{1}}{2}\right]  .\label{55}%
\end{align}
Performing integration over the phase space (using arguments of the $f$
function as integration variables in each term) one arrives at the following
excitation probability:%
\begin{subequations}%
\label{56}%
\begin{align}
w &  =w_{bg}+\tilde{w}\int d\mathbf{r}d\mathbf{p}\cos\left(  \phi_{m}-\phi
_{R}-\phi_{4}+2\phi_{3}-2\phi_{2}+\phi_{1}\right)  f\left(  \mathbf{r}%
,\mathbf{p}\right)  ,\label{56a}\\
w_{bg} &  =\tilde{w}=\dfrac{1}{2},\label{56b}\\
\phi_{m} &  =\mathbf{k}_{4}\cdot\mathbf{R}\left(  \mathbf{r},\mathbf{p}%
+\dfrac{\hbar\mathbf{k}_{1}}{2},T_{4}\right)  -2\mathbf{k}_{3}\cdot
\mathbf{R}\left(  \mathbf{r},\mathbf{p}+\dfrac{\hbar\mathbf{k}_{1}}{2}%
,T_{3}\right)  +2\mathbf{k}_{2}\cdot\mathbf{R}\left(  \mathbf{r}%
,\mathbf{p}+\dfrac{\hbar\mathbf{k}_{1}}{2},T_{2}\right)  -\mathbf{k}_{1}%
\cdot\mathbf{r},\label{56c}\\
\phi_{R} &  =\delta_{12}\left(  T_{4}-2T_{3}+2T_{2}\right)  ,\label{56d}%
\end{align}%
\end{subequations}%
where $w_{bg}$ and $\tilde{w}$ are the background and the amplitude of the
interference term correspondingly. Interferometer phase (\ref{56c}) coincides
with that given in Eq. (\ref{2}) if all pulses have the same effective wave
vectors, $\mathbf{k}_{i}=\mathbf{k}$.

For arbitrary areas of pulses one gets 34 terms, among them are the following:

\begin{itemize}
\item Eight terms correspond to the transfer of level populations between
fields, they comprise background of the excitation probability given by%
\begin{equation}
w_{bg}=\dfrac{1}{2}\left(  1-\cos\theta_{1}\cos\theta_{2}\cos\theta_{3}%
\cos\theta_{4}\right)  .\label{57}%
\end{equation}

\item 17 Ramsey terms, which are washed out after averaging over momenta.

\item Six terms correspond to the coherence transfer between two adjacent time
intervals. They have structure similar to the case of the excitation by three
pulses and are also washed out at least because the ratio of time intervals
does not satisfy echo conditions.

\item Two terms $w_{\pm}$ associated with the stimulated echo. They are given
by%
\begin{align}
w_{\pm} &  =-.25\sin\theta_{4}\sin\theta_{3}\sin\theta_{2}\sin\theta_{1}\int
d\mathbf{r}d\mathbf{p}\nonumber\\
&  \times\cos\left\{  \mathbf{k}_{4}\cdot\mathbf{R}\left[  \mathbf{R}\left(
\mathbf{R}\left(  \mathbf{r},\mathbf{p}+\dfrac{\hbar\mathbf{k}_{1}}{2}%
,T_{2}\right)  ,\mathbf{P}\left(  \mathbf{r},\mathbf{p}+\dfrac{\hbar
\mathbf{k}_{1}}{2},T_{2}\right)  \pm\dfrac{\hbar\mathbf{k}_{2}}{2},T_{3}%
-T_{2}\right)  ,\right.  \right.  \nonumber\\
&  \left.  \mathbf{P}\left(  \mathbf{R}\left(  \mathbf{r},\mathbf{p}%
+\dfrac{\hbar\mathbf{k}_{1}}{2},T_{2}\right)  ,\mathbf{P}\left(
\mathbf{r},\mathbf{p}+\dfrac{\hbar\mathbf{k}_{1}}{2},T_{2}\right)  \pm
\dfrac{\hbar\mathbf{k}_{2}}{2},T_{3}-T_{2}\right)  \mp\dfrac{\hbar
\mathbf{k}_{3}}{2},T_{4}-T_{3}\right]  \nonumber\\
&  \left.  -\mathbf{k}_{3}\cdot\mathbf{R}\left[  \mathbf{R}\left(
\mathbf{r},\mathbf{p}+\dfrac{\hbar\mathbf{k}_{1}}{2},T_{2}\right)
,\mathbf{P}\left(  \mathbf{r},\mathbf{p}+\dfrac{\hbar\mathbf{k}_{1}}{2}%
,T_{2}\right)  +\dfrac{\hbar\mathbf{k}_{2}}{2},T_{3}-T_{2}\right]
-\delta_{12}\left(  T_{4}-T_{3}\right)  -\phi_{4}+\phi_{3}\right\}
\nonumber\\
&  \times\cos\left[  \mathbf{k}_{2}\cdot\mathbf{R}\left(  \mathbf{r}%
,\mathbf{p}+\dfrac{\hbar\mathbf{k}_{1}}{2},T_{2}\right)  -\mathbf{k}_{1}%
\cdot\mathbf{r}-\delta_{12}T_{2}-\phi_{2}+\phi_{1}\right]  f\left(
\mathbf{r},\mathbf{p}\right)  ;\label{58}%
\end{align}

\item A term associated with the double-loop interferometer. It coincides with
the second term in Eq. (\ref{56a}), except that the amplitude is equal to%
\begin{equation}
\tilde{w}=\dfrac{1}{2}\sin\theta_{4}\sin^{2}\dfrac{\theta_{3}}{2}\sin
^{2}\dfrac{\theta_{2}}{2}\sin\theta_{1}.\label{59}%
\end{equation}

\end{itemize}

\subsection{\label{s7a}Atomic gyroscope}

This is evident from Eq. (\ref{56c}) that, for an arbitrary rotation and
acceleration, one can eliminate phase shifts of the order $T_{j}^{n}$ only if
wave vectors and ratios of time intervals (\ref{401}) satisfy the equality%
\begin{equation}
\mathbf{k}_{4}-2\mathbf{k}_{3}t_{3}^{n}+2\mathbf{k}_{2}t_{2}^{n}%
-\mathbf{k}_{1}t_{1}^{n}=0.\label{60}%
\end{equation}
We always need to eliminate zero-order terms [phase matching condition
(\ref{1a})] and first-order terms (to get interference). For the gyroscope one
needs also to eliminate the second-order term. From the system of Eqs.
(\ref{60}) with $n=0,$ $1,$ and $2$ one concludes that all effective wave
vectors have to be collinear,%
\begin{equation}
\mathbf{k}_{j}=s_{j}\mathbf{k,}\label{61}%
\end{equation}
where $\mathbf{k}$ is comprised of counterpropagating traveling waves,
$s_{j}=\sin\alpha_{j},$ and $\alpha_{j}$ is one-half of the angle between
optical fields wave vectors $\mathbf{q}_{1j}$ and $\mathbf{q}_{2j}$ presented
in the pulse $j$. Resolving the set of equations, for example, with respect to
variables $s_{1},s_{2},$ and $s_{3},$ one finds that%
\begin{subequations}%
\label{62}%
\begin{gather}
s_{1}=s_{4}\dfrac{\left(  1-\ t_{2}\right)  \left(  1-\ t_{3}\right)
}{\ t_{2}\ t_{3}},\label{62a}\\
s_{2}=\dfrac{s_{4}}{2}\dfrac{1-\ t_{3}}{\ t_{2}\left(  \ t_{3}-\ t_{2}\right)
},\label{62b}\\
s_{3}=\dfrac{s_{4}}{2}\dfrac{1-\ t_{2}}{\ t_{3}\left(  \ t_{3}-\ t_{2}\right)
}.\label{62c}%
\end{gather}%
\end{subequations}%
One sees that for given \ time delays between pulses one can always choose
proper values of effective wave vectors to construct a double-loop
interferometer. Moreover, wave vectors can be scaled by a given factor. Since
the interferometer phase (\ref{56c}) is linear with respect to wave vectors
and, therefore, proportional to this scale factor, one can choose it from
conditions%
\begin{equation}
s_{j}\leq1,\label{63}%
\end{equation}
i.e.,%
\begin{equation}
s_{4}=\min\left[  1,\dfrac{\ t_{2}\ t_{3}}{\left(  1-\ t_{2}\right)  \left(
1-\ t_{3}\right)  },\dfrac{2\ t_{2}\left(  \ t_{3}-\ t_{2}\right)  }%
{1-\ t_{3}},\dfrac{2\ t_{3}\left(  \ t_{3}-\ t_{2}\right)  }{1-\ t_{2}%
}\right]  .\label{64}%
\end{equation}
This expression guaranteed that all wave vectors satisfy the condition
(\ref{63}) and at least one of them yields the maximum value $k$ (or $s_{j}=1$).

\begin{figure} [htbp]
\vspace{0.5cm}
\includegraphics[height=12.0cm,width=12.0cm]{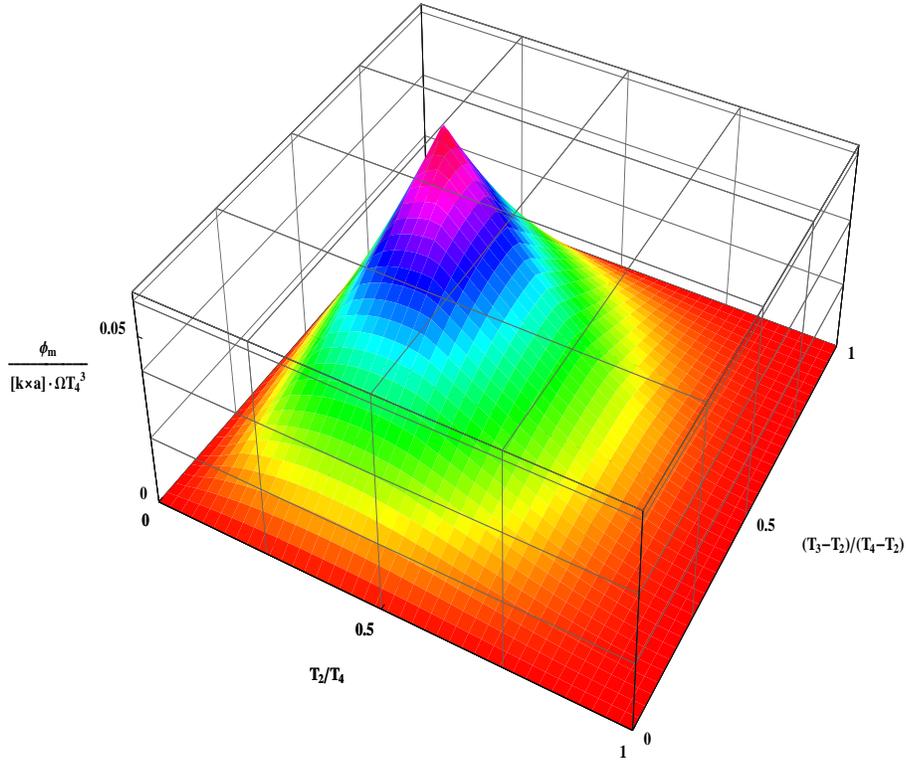}
\caption{Dependence of the gyroscope phase on the position of the second and
third pulse.}
\label{f2}
\end{figure}

To get phase shifts of the atomic gyroscope, one can simply notice from
comparison of Eqs. (\ref{51b}) and (\ref{56c}) that it is sufficient to
replace $T_{3}^{n}$ in column 1 of Table \ref{t1} by%
\begin{equation}
T_{3}^{n}\rightarrow T_{4}^{n}\dfrac{s_{4}-2s_{3}t_{3}^{n}+2s_{2}t_{2}^{n}%
}{1-2^{1-n}}.\label{65}%
\end{equation}
Since all $T^{2}$ terms are eliminated, for the launch momentum%
\[
p\ll ma/\Omega
\]
the main phase shift is presented in the fourth row of Table \ref{t1}. Saving
only this term after the replacement (\ref{65}) one finds%
\begin{subequations}%
\label{66}%
\begin{align}
\phi_{m} &  =\mathbf{\Omega}\cdot\left(  \mathbf{k}\times\mathbf{a}\right)
T_{4}^{3}\psi,\label{66a}\\
\psi &  =\dfrac{s_{4}}{3}\left(  1-\ t_{2}\right)  \left(  1-\ t_{3}\right)
.\label{66b}%
\end{align}%
\end{subequations}
Phase dependence (\ref{66}) on the relative position of the second and the
third pulses (on the parameters $t_{2}\ $and $t_{3})$ for the optimal value
(\ref{64}) of the parameter $s_{4}$ is plotted in Fig. \ref{f2}.

One can see that the phase (\ref{66}) has only one maximum and that this
maximum occurs for the double-loop interferometer shown in Fig. \ref{f1}(a),
when
\begin{equation}
s_{j}=1~\text{and~}t_{2}=1/4,t_{3}=3/4,\label{67}%
\end{equation}
and $\phi_{m}$ is given by Eq. (\ref{6}).

Nevertheless, one has to move out of the point (\ref{67}) at least slightly to
wash out contributions from the stimulated echo (\ref{58}). Let us consider
this case. If
\begin{equation}
s_{j}=1-\sigma_{j}\label{671}%
\end{equation}
and $\sigma_{j}\ll1,$ then from Eqs. (\ref{62}) one finds with accuracy up to
second order terms in $\sigma_{j},$ that the second and the third pulses have
to be located at%
\begin{subequations}%
\label{672}
\begin{align}
t_{2} &  \approx\dfrac{1}{4}\left(  1+\dfrac{9}{8}\sigma_{1}-\sigma_{2}%
-\dfrac{1}{8}\sigma_{4}+\dfrac{3}{4}\sigma_{1}\sigma_{2}-\dfrac{3}{16}%
\sigma_{1}\sigma_{4}+\dfrac{1}{4}\sigma_{2}\sigma_{4}+\dfrac{9}{32}\sigma
_{1}^{2}-\sigma_{2}^{2}-\dfrac{3}{32}\sigma_{4}^{2}\right)  ,\label{672a}\\
t_{3} &  \approx\dfrac{1}{4}\left(  3-\dfrac{3}{8}\sigma_{1}+\sigma_{2}%
-\dfrac{5}{8}\sigma_{4}+\dfrac{1}{4}\sigma_{2}\sigma_{4}-\dfrac{1}{16}%
\sigma_{1}\sigma_{4}-\dfrac{5}{4}\sigma_{1}\sigma_{2}+\dfrac{15}{32}\sigma
_{1}^{2}+\sigma_{2}^{2}-\dfrac{13}{32}\sigma_{4}^{2}\right)  .\label{672b}%
\end{align}%
\end{subequations}%
We found these values from Eqs. (\ref{62a}) and (\ref{62b}). Substituting them
into Eq. (\ref{62c}) one gets the phase matching condition, in the first order
over $\sigma_{j},$
\begin{equation}
\sigma_{4}-2\sigma_{3}+2\sigma_{2}-\sigma_{1}=0,\label{68}%
\end{equation}
and the equality in the second order. We calculate, with the same accuracy,
the main term (\ref{66}) in the gyroscope phase,%
\begin{equation}
\psi\approx\dfrac{1}{16}-\dfrac{1}{48}\sigma_{4}-\dfrac{1}{24}\sigma
_{2}-\allowbreak\dfrac{45}{1024}\sigma_{1}^{2}-\dfrac{1}{16}\sigma_{2}%
^{2}-\dfrac{13}{1024}\sigma_{4}^{2}+\dfrac{3}{32}\sigma_{1}\sigma_{2}%
-\dfrac{3}{512}\sigma_{1}\sigma_{4}+\dfrac{1}{32}\sigma_{2}\sigma
_{4},\label{69}%
\end{equation}
while all terms in the zero-order over $\sigma_{j},$ calculated using Table
\ref{t1} and the rule (\ref{65}), are presented in Table \ref{t2}. In the last
two columns of this table and of Table \ref{t3} below we calculated phase for
Cs interferometer ($m=2.21\times10^{-22}~$g, $k=1.47\times10^{5}%
~$cm$^{\text{-1}}$) for $g_{z}=-980~cm/s^{2},$ $T_{4}=0.8~s,$ $R=6.72\times
10^{8}$ cm, latitude 41$^{\circ}$.

\begin{table}[ptb]
\caption{Atomic gyroscope phase shifts; column1,  the general case; column 2,
for the wave vector $\mathbf{k}$ and launch momentum $\mathbf{p}$ in the
west-east and vertical directions; columns 3 and 4 represent the absolute and
the relative phase shifts when $p=0$ for Cs gyroscope }%
\begin{tabular}
[c]{l}\hline\hline
$%
\begin{array}
[c]{ccccccc}%
\smallskip\dfrac{3\Omega^{2}}{32m}\left[  \left(  \mathbf{k}\boldsymbol{\nu
}\right)  \left(  \mathbf{p}\boldsymbol{\nu}\right)  -\left(  \mathbf{kp}%
\right)  \right]  T_{4}^{3} & \hspace{0.5in} & 0 & \hspace{0.5in} & 0 &
\hspace{0.5in} & 0\\
\smallskip\dfrac{1}{16}\boldsymbol{\Omega}\cdot\left(  \mathbf{k}%
\times\mathbf{a}\right)  T_{4}^{3} &  &
\begin{array}
[c]{c}%
\smallskip-\dfrac{1}{16}k\nu_{y}g_{z}\Omega T_{4}^{3}\\
-\dfrac{1}{16}k\nu_{y}\Omega^{3}RT_{4}^{3}%
\end{array}
&  &
\begin{array}
[c]{c}%
\smallskip254\\
0.92
\end{array}
&  &
\begin{array}
[c]{c}%
\smallskip1\\
3.6\times10^{-3}%
\end{array}
\\
\smallskip\dfrac{3\hbar}{64m}\left[  \left(  \mathbf{k}\cdot\boldsymbol{\nu
}\right)  ^{2}-k^{2}\right]  \Omega^{2}T_{4}^{3} &  & -\dfrac{3\hbar k^{2}%
}{64m}\Omega^{2}T_{4}^{3} &  & -1.32\times10^{-5} &  & -5.2\times10^{-8}\\
\smallskip-\dfrac{3}{64}\left[  \left(  \mathbf{k}\cdot\mathbf{a}\right)
-\left(  \mathbf{k}\cdot\boldsymbol{\nu}\right)  \left(  \mathbf{a}%
\cdot\boldsymbol{\nu}\right)  \right]  \Omega^{2}T_{4}^{4} &  & 0 &  & 0 &  &
0\\
\smallskip-\dfrac{1}{16m}\left(  \boldsymbol{\nu}\times\mathbf{k}\right)
\cdot\mathbf{p}\Omega^{3}T_{4}^{4} &  & \dfrac{\nu_{y}kp}{16m}\Omega^{3}%
T_{4}^{4} &  & 0 &  & 0\\
\smallskip-\dfrac{1}{64m}\mathbf{k}\cdot\mathbf{p}T_{4}^{3}T_{zz} &  & 0 &  &
0 &  & 0\\
\smallskip-\dfrac{\hbar k^{2}}{128m}T_{4}^{3}T_{zz} &  & -\dfrac{\hbar k^{2}%
}{128m}T_{4}^{3}T_{zz} &  & -1.21\times10^{-3} &  & 4.8\times10^{-6}\\
\smallskip-\dfrac{1}{128}\mathbf{k}\cdot\mathbf{a}T_{4}^{4}T_{zz} &  & 0 &  &
0 &  & 0\\
\smallskip\dfrac{3}{64m}\left(  \mathbf{n}\cdot\boldsymbol{\nu}\right)
\left(  \mathbf{k}\cdot\mathbf{n}\right)  \left(  \boldsymbol{\nu}%
\cdot\mathbf{p}\right)  T_{4}^{3}T_{zz} &  & 0 &  & 0 &  & 0\\
\smallskip\dfrac{3\hbar}{128m}\left(  \mathbf{n}\cdot\boldsymbol{\nu}\right)
\left(  \mathbf{k}\cdot\mathbf{n}\right)  \left(  \boldsymbol{\nu}%
\cdot\mathbf{k}\right)  T_{4}^{3}T_{zz} &  & 0 &  & 0 &  & 0\\
\smallskip\dfrac{3}{128}\left(  \mathbf{n}\cdot\boldsymbol{\nu}\right)
\left(  \mathbf{k}\cdot\mathbf{n}\right)  \left(  \boldsymbol{\nu}%
\cdot\mathbf{a}\right)  T_{4}^{4}T_{zz} &  & 0 &  & 0 &  & 0\\
\smallskip-\dfrac{3}{64m}\left(  \mathbf{n}\cdot\boldsymbol{\nu}\right)
\left[  \boldsymbol{\nu}\cdot\left(  \mathbf{n}\times\mathbf{k}\right)
\right]  \left(  \boldsymbol{\nu}\cdot\mathbf{p}\right)  \Omega T_{4}%
^{4}T_{zz} &  & -\dfrac{3kp}{64m}\nu_{y}\nu_{z}^{2}\Omega T_{4}^{4}T_{zz} &  &
0 &  & 0\\
\smallskip-\dfrac{3\hbar}{128m}\left(  \mathbf{n}\cdot\boldsymbol{\nu}\right)
\left[  \boldsymbol{\nu}\cdot\left(  \mathbf{n}\times\mathbf{k}\right)
\right]  \left(  \boldsymbol{\nu}\cdot\mathbf{k}\right)  \Omega T_{4}%
^{4}T_{zz} &  & 0 &  & 0 &  & 0\\
\smallskip\dfrac{3}{64m}\left[  \left(  \boldsymbol{\nu}\times\mathbf{n}%
\right)  \cdot\left(  \boldsymbol{\nu}\times\mathbf{p}\right)  \right]
\left(  \mathbf{k}\cdot\mathbf{n}\right)  T_{4}^{3}T_{zz} &  & 0 &  & 0 &  &
0\\
\smallskip\dfrac{3\hbar}{128m}\left[  \left(  \boldsymbol{\nu}\times
\mathbf{n}\right)  \cdot\left(  \boldsymbol{\nu}\times\mathbf{k}\right)
\right]  \left(  \mathbf{k}\cdot\mathbf{n}\right)  T_{4}^{3}T_{zz} &  & 0 &  &
0 &  & 0\\
\smallskip-\dfrac{3}{64m}\left[  \left(  \boldsymbol{\nu}\times\mathbf{n}%
\right)  \cdot\left(  \boldsymbol{\nu}\times\mathbf{p}\right)  \right]
\left[  \boldsymbol{\nu}\cdot\left(  \mathbf{n}\times\mathbf{k}\right)
\right]  \Omega T_{4}^{4}T_{zz} &  & -\dfrac{3kp\nu_{y}^{3}}{64m}\Omega
T_{4}^{4}T_{zz} &  & 0 &  & 0\\
\smallskip-\dfrac{3\hbar}{128m}\left[  \left(  \boldsymbol{\nu}\times
\mathbf{n}\right)  \cdot\left(  \boldsymbol{\nu}\times\mathbf{k}\right)
\right]  \left[  \boldsymbol{\nu}\cdot\left(  \mathbf{n}\times\mathbf{k}%
\right)  \right]  \Omega T_{4}^{4}T_{zz} &  & 0 &  & 0 &  & 0\\
\smallskip\dfrac{3}{64m}\left[  \left(  \boldsymbol{\nu}\times\mathbf{n}%
\right)  \cdot\mathbf{p}\right]  \left(  \mathbf{k}\cdot\mathbf{n}\right)
\Omega T_{4}^{4}T_{zz} &  & 0 &  & 0 &  & 0\\
\smallskip\dfrac{3\hbar}{128m}\left[  \left(  \boldsymbol{\nu}\times
\mathbf{n}\right)  \cdot\mathbf{k}\right]  \left(  \mathbf{k}\cdot
\mathbf{n}\right)  \Omega T_{4}^{4}T_{zz} &  & 0 &  & 0 &  & 0\\
\smallskip\dfrac{3}{128}\left[  \left(  \boldsymbol{\nu}\times\mathbf{n}%
\right)  \cdot\left(  \boldsymbol{\nu}\times\mathbf{a}\right)  \right]
\left(  \mathbf{k}\cdot\mathbf{n}\right)  T_{4}^{4}T_{zz} &  & 0 &  & 0 &  & 0
\end{array}
$\\\hline\hline
\end{tabular}
\label{t2}%
\end{table}

\subsection{\label{s7b}Atomic accelerometer}

\begin{figure} [htbp]
\vspace{0.5cm}
\includegraphics[height=12.0cm,width=12.0cm]{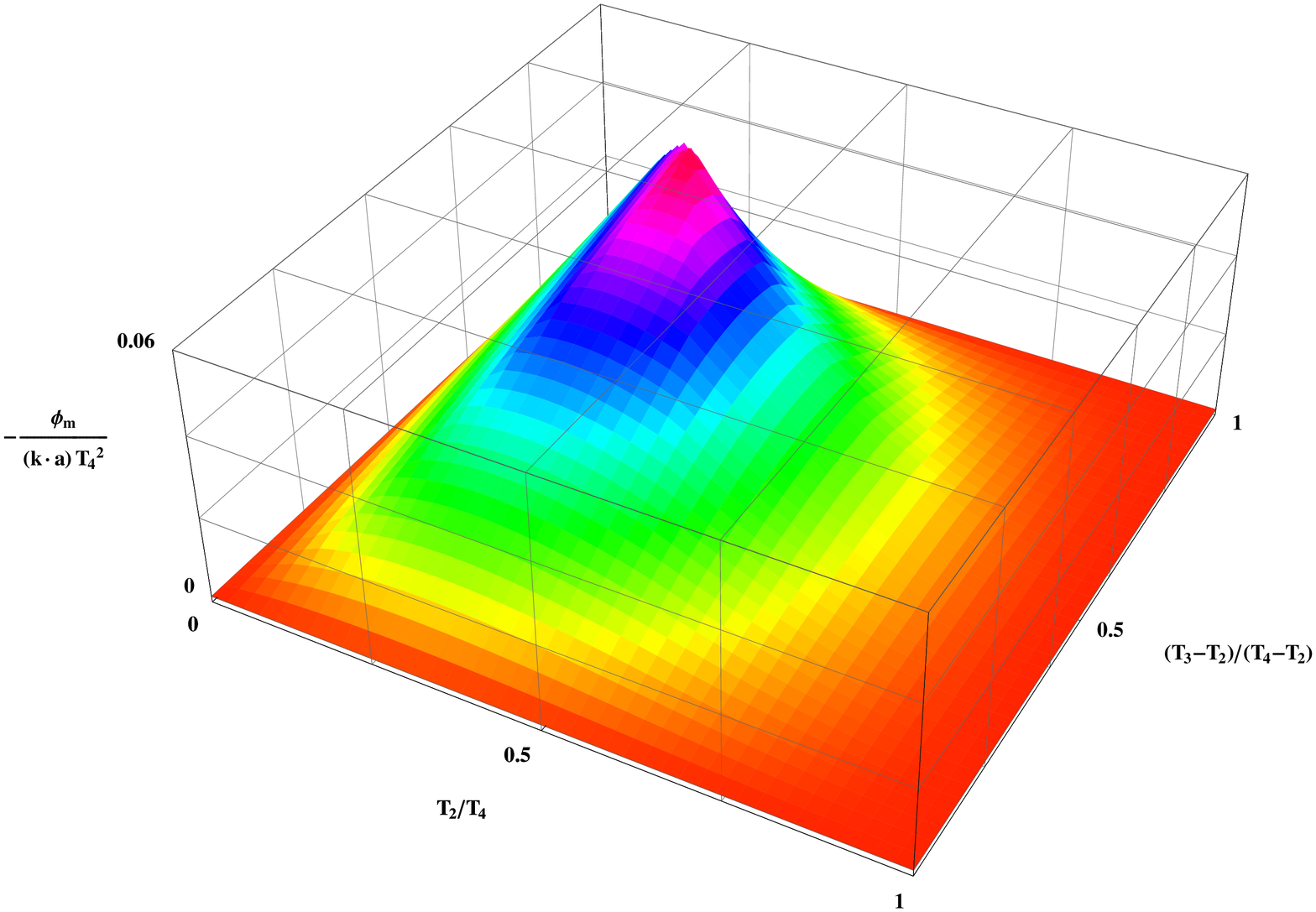}
\caption{The same as in Fig. \ref{f2}, but for the phase of accelerometer.}
\label{f3}
\end{figure}

One can use the four-pulse interferometer to eliminate third-order
contributions to the interferometer phase in order to increase the accuracy of
acceleration measurements. In this case wave vectors and relative positions of
the second and the third pulses have to satisfy Eqs. (\ref{60}) for $n=0,~1,$
and $3.$ Resolving these equations with respect to wave vectors of the first,
second, and third pulses and requiring that at least one of the pulse is
comprised of counterpropagating waves, one finds%
\begin{subequations}%
\label{70}%
\begin{align}
s_{1} &  =s_{4}\dfrac{\left(  1-t_{3}\right)  \left(  1-t_{2}\right)  \left(
1+t_{2}+t_{3}\right)  }{t_{2}t_{3}\left(  t_{2}+t_{3}\right)  },\label{70a}\\
s_{2} &  =\dfrac{s_{4}}{2}\dfrac{1-t_{3}^{2}}{t_{2}\left(  t_{3}^{2}-t_{2}%
^{2}\right)  },\label{70b}\\
s_{3} &  =\dfrac{s_{4}}{2}\dfrac{1-t_{2}^{2}}{t_{3}\left(  t_{3}^{2}-t_{2}%
^{2}\right)  },\label{70c}\\
s_{4} &  =\min\left[  \dfrac{t_{2}t_{3}\left(  t_{2}+t_{3}\right)  }{\left(
1-t_{3}\right)  \left(  1-t_{2}\right)  \left(  1+t_{2}+t_{3}\right)  }%
,\dfrac{2t_{2}\left(  t_{3}^{2}-t_{2}^{2}\right)  }{1-t_{3}^{2}},\dfrac
{2t_{3}\left(  t_{3}^{2}-t_{2}^{2}\right)  }{1-t_{2}^{2}},1\right]
.\label{70d}%
\end{align}%
\end{subequations}%
To get the main contribution to the interferometer phase one should perform
the replacement (\ref{65}) for $n=2$ in the very first cell of Table \ref{t1}.
It yields the following expression:%
\begin{equation}
\phi_{m}=-\dfrac{s_{4}\left(  1-t_{3}\right)  \left(  1-t_{2}\right)
}{\allowbreak2\left(  t_{2}+t_{3}\right)  }\mathbf{k}\cdot\mathbf{a}T_{4}%
^{2}.\label{71}%
\end{equation}
This dependence is shown in Fig. \ref{f3}.

One sees that only one maximum of the accelerometer's phase corresponds to
values of $t_{2}$ and $t_{3}$ given by Eq. (\ref{8}). Substituting these
values into Eqs. (\ref{70}) one verifies that the maximum of accelerometer
phase occurs when all pulses are comprised of counterpropagating fields,
$s_{j}=1.$

From Eq. (\ref{8}) one can see that $T_{4}-T_{3}\not =T_{2}$ and therefore the
contribution from stimulated echo processes given by Eq. (\ref{58}) is washed
out after averaging over momenta.

Different terms in the accelerometer phase can be obtained from Table \ref{t1}
using the replacement (\ref{65}). They are presented in Table \ref{t3}.

\begin{table}[ptb]
\caption{The same as in Table \ref{t1}, but for the four-pulse interferometer
shown in Fig. \ref{f1}(c). The last two columns are for the phase shifts'
numerical\ and relative values of the Cs interferometer for ${}p=0$.}%
\begin{tabular}
[c]{l}\hline\hline
$%
\begin{array}
[c]{ccccc}%
\smallskip-\dfrac{\sqrt{5}-2}{4}\mathbf{k}\cdot\mathbf{a}T_{4}^{2} &
\begin{array}
[c]{c}%
\smallskip-\dfrac{\sqrt{5}-2}{4}kg_{z}T_{4}^{2}\\
-\dfrac{\sqrt{5}-2}{4}k\Omega_{y}^{2}RT_{4}^{2}%
\end{array}
&  &
\begin{array}
[c]{c}%
\smallskip5.46\times10^{6}\\
-1.13\times10^{4}%
\end{array}
&
\begin{array}
[c]{c}%
1\\
2.0\times10^{-3}%
\end{array}
\\
\smallskip-\dfrac{\sqrt{5}-2}{2m}\left(  \boldsymbol{\Omega}\times
\mathbf{k}\right)  \cdot\mathbf{p}T_{4}^{2} & 0 &  & 0 & 0\\
\smallskip-\dfrac{8-3\sqrt{5}}{64}\left[  \left(  \mathbf{k}\cdot
\mathbf{a}\right)  -\left(  \mathbf{k}\cdot\boldsymbol{\nu}\right)  \left(
\mathbf{a}\cdot\boldsymbol{\nu}\right)  \right]  \Omega^{2}T_{4}^{4} &
\begin{array}
[c]{c}%
\smallskip-\dfrac{8-3\sqrt{5}}{64}kg_{z}\Omega_{y}^{2}T_{4}^{4}\\
-\dfrac{8-3\sqrt{5}}{64}k\Omega^{2}\Omega_{y}^{2}RT_{4}^{4}%
\end{array}
&  &
\begin{array}
[c]{c}%
\smallskip3.60\times10^{-3}\\
1.31\times10^{-5}%
\end{array}
&
\begin{array}
[c]{c}%
6.6\times10^{-10}\\
2.4\times10^{-12}%
\end{array}
\\
\smallskip-\dfrac{8-3\sqrt{5}}{48m}\left(  \boldsymbol{\nu}\times
\mathbf{k}\right)  \cdot\mathbf{p}\Omega^{3}T_{4}^{4} & 0 &  &  & \\
\smallskip-\dfrac{8-3\sqrt{5}}{384}\mathbf{k}\cdot\mathbf{a}T_{4}^{4}T_{zz} &
\begin{array}
[c]{c}%
\smallskip\underline{-\dfrac{8-3\sqrt{5}}{384}kg_{z}T_{3}^{4}T_{zz}}\\
\underleftarrow{-\dfrac{8-3\sqrt{5}}{384}k\Omega_{y}^{2}RT_{4}^{4}T_{zz}}%
\end{array}
&
\begin{array}
[c]{c}%
\dfrac{8-3\sqrt{5}}{192}kg_{z}T_{4}^{4}T_{zz}\\
\dfrac{8-3\sqrt{5}}{192}k\Omega_{y}^{2}RT_{4}^{4}T_{zz}%
\end{array}
&
\begin{array}
[c]{c}%
\smallskip-1.16\\
2.40\times10^{-3}%
\end{array}
&
\begin{array}
[c]{c}%
2.1\times10^{-7}\\
4.4\times10^{-10}%
\end{array}
\\
\smallskip\dfrac{8-3\sqrt{5}}{128}\left(  \mathbf{n}\cdot\boldsymbol{\nu
}\right)  \left(  \mathbf{k}\cdot\mathbf{n}\right)  \left(  \boldsymbol{\nu
}\cdot\mathbf{a}\right)  T_{4}^{4}T_{zz} & \underline{\dfrac{8-3\sqrt{5}}%
{128}\nu_{z}^{2}kg_{z}T_{3}^{4}T_{zz}} &  &  & \\
\smallskip-\dfrac{8-3\sqrt{5}}{64m}\left(  \mathbf{n}\cdot\boldsymbol{\nu
}\right)  \left[  \boldsymbol{\nu}\cdot\left(  \mathbf{n}\times\mathbf{k}%
\right)  \right]  \left(  \boldsymbol{\nu}\cdot\mathbf{p}\right)  \Omega
T_{4}^{4}T_{zz} & 0 &  &  & \\
\smallskip-\dfrac{8-3\sqrt{5}\hbar}{128m}\left(  \mathbf{n}\cdot
\boldsymbol{\nu}\right)  \left[  \boldsymbol{\nu}\cdot\left(  \mathbf{n}%
\times\mathbf{k}\right)  \right]  \left(  \boldsymbol{\nu}\cdot\mathbf{k}%
\right)  \Omega T_{4}^{4}T_{zz} & 0 &  &  & \\
\smallskip-\dfrac{8-3\sqrt{5}}{64m}\left[  \left(  \boldsymbol{\nu}%
\times\mathbf{n}\right)  \cdot\left(  \boldsymbol{\nu}\times\mathbf{p}\right)
\right]  \left[  \boldsymbol{\nu}\cdot\left(  \mathbf{n}\times\mathbf{k}%
\right)  \right]  \Omega T_{4}^{4}T_{zz} & 0 &  &  & \\
\smallskip-\dfrac{\left(  8-3\sqrt{5}\right)  \hbar}{128m}\left[  \left(
\boldsymbol{\nu}\times\mathbf{n}\right)  \cdot\left(  \boldsymbol{\nu}%
\times\mathbf{k}\right)  \right]  \left[  \boldsymbol{\nu}\cdot\left(
\mathbf{n}\times\mathbf{k}\right)  \right]  \Omega T_{4}^{4}T_{zz} & 0 &  &  &
\\
\smallskip\dfrac{8-3\sqrt{5}}{64m}\left[  \left(  \boldsymbol{\nu}%
\times\mathbf{n}\right)  \cdot\mathbf{p}\right]  \left(  \mathbf{k}%
\cdot\mathbf{n}\right)  \Omega T_{4}^{4}T_{zz} & 0 &  &  & \\
\smallskip\dfrac{\left(  8-3\sqrt{5}\right)  \hbar}{128m}\left[  \left(
\boldsymbol{\nu}\times\mathbf{n}\right)  \cdot\mathbf{k}\right]  \left(
\mathbf{k}\cdot\mathbf{n}\right)  \Omega T_{4}^{4}T_{zz} & 0 &  &  & \\
\smallskip\dfrac{8-3\sqrt{5}}{128}\left[  \left(  \boldsymbol{\nu}%
\times\mathbf{n}\right)  \cdot\left(  \boldsymbol{\nu}\times\mathbf{a}\right)
\right]  \left(  \mathbf{k}\cdot\mathbf{n}\right)  T_{4}^{4}T_{zz} &
\begin{array}
[c]{c}%
\smallskip\underline{\dfrac{8-3\sqrt{5}}{128}k\nu_{y}^{2}g_{z}T_{3}^{4}T_{zz}%
}\\
\underleftarrow{\dfrac{8-3\sqrt{5}}{128}k\nu_{y}^{2}\Omega^{2}RT_{4}^{4}%
T_{zz}}%
\end{array}
&  &  &
\end{array}
$\\\hline\hline
\end{tabular}
\label{t3}%
\end{table}

\section{\label{s8}Discussion}

In this paper we showed that, in the presence of inertial forces and
homogeneous acceleration gradient, the density matrix in the Wigner
representation in the free space still obeys classical Liouville equation for
the distribution function. It allows us to get interferometers' phases without
calculating path integrals.

For all interferometers under consideration we calculate phase shifts up to
terms proportional to $T^{4}$ for an arbitrary constant in time acceleration
and rotation.

Using double-loop interferometers $\pi/2-\pi-\pi-\pi/2$ one can eliminate the
phase dependence on the acceleration only and get an interferometer with the
main phase term proportional to the cross product of acceleration and rotation
frequency. One can use this interferometer as a gyroscope to measure rotation
frequency components perpendicular to the acceleration. Even for the rotation
frequency as small as the Earth rotation rate, the phase of this gyroscope can
achieve $\phi_{m}\sim100.$ Eliminating the acceleration term, which is
typically five orders of magnitude larger, could be incomplete if pulse areas
are not precisely equal to $\pi/2$ or $\pi,$ because in this case stimulated
echo processes, which are extremely sensitive to the acceleration, contribute
to the probability of the particles' excitation. To exclude these processes we
propose to drive Raman transitions between hyperfine atomic sublevels by a
pulse of travelling noncounterpropagating optical waves. For this goal one can
use even slightly noncounterpropagating fields having effective wave vectors
given by Eqs. (\ref{61}) and (\ref{671}), where the small parameter
$\sigma_{j}$ determines the deviation of the pulse $j$ wave vector from the
maximum value $k.$ If wave vectors satisfy the phase matching condition
(\ref{1a}) for double-loop interferometer but violate the phase-matching
condition (\ref{1b}) for stimulated echo, i.e., if%
\begin{subequations}%
\label{72}%
\begin{align}
\sigma_{dl} &  =\sigma_{4}-2\sigma_{3}+2\sigma_{2}-\sigma_{1}=0,\label{72a}\\
\sigma_{st} &  =\sigma_{4}-\sigma_{3}-\sigma_{2}+\sigma_{1}\not =0,\label{72b}%
\end{align}%
\end{subequations}%
integrands in Eq. (\ref{58}) for probabilities of excitation associated with
stimulated echo processes become rapidly oscillating in space function, which
have the period of the order of$\ \lambda/\sigma_{st},$ where $\lambda=2\pi/k$
is the effective wavelength. When
\begin{equation}
\dfrac{\lambda}{s_{cloud}}\ll\sigma_{st}\ll1,\label{73}%
\end{equation}
where $s_{cloud}$ is the initial atomic cloud size, in the case when the
initial atom density is smooth in space, probabilities (\ref{58}) become
exponentially small with respect to the parameter $\sigma_{st}s_{cloud}%
/\lambda\gg1.$ In addition to arguments related to the difference in phase
matching conditions, one can expect further decrease of probabilities
(\ref{58}) due to the fact that pulses are not properly positioned in time
[$1-t_{3}\not =t2,$ where $t_{j}$ are given by Eqs. (\ref{672})] to produce
the stimulated echo. We obtained the required pulses' positions and the main
term in the gyroscope phase for arbitrary $\sigma_{j},$ see Eqs. (\ref{672}),
(\ref{66a}), and (\ref{69}). But it is sufficient for only two effective wave
vectors to have $\sigma_{j}\not =0.$ If, for example,%
\begin{equation}
\sigma_{3}=\sigma_{4}=0~\text{and }\sigma_{2}=\dfrac{\sigma_{1}}{2}%
\not =0,\label{74}%
\end{equation}
then $\sigma_{st}=\dfrac{\sigma_{1}}{2},$ second and third pulses have to be
applied at moments%
\begin{subequations}%
\label{75}%
\begin{align}
T_{2} &  \approx\dfrac{T_{4}}{4}\left(  1+\dfrac{5}{8}\sigma_{1}+\dfrac
{13}{32}\sigma_{1}^{2}\right)  ,\label{75a}\\
T_{3} &  \approx\dfrac{3}{4}T_{4}\left(  1+\dfrac{1}{24}\sigma_{1}+\dfrac
{1}{32}\sigma_{1}^{2}\right)  \label{75b}%
\end{align}%
\end{subequations}%
and gyroscope phase is given by%
\begin{equation}
\phi_{m}=\dfrac{1}{16}\left(  1-\dfrac{1}{3}\sigma_{1}-\dfrac{13}{64}%
\sigma_{1}^{2}\right)  \left(  \mathbf{k}\times\mathbf{a}\right)
\cdot\mathbf{\Omega}T_{4}^{3}.\label{76}%
\end{equation}

A \textquotedblleft side-effect" of using Raman pulses with nonequal effective
wave vectors is that in this case one gets Ramsey fringes, i.e., excitation
probability oscillating dependences on the Raman detuning $\delta_{12}$ having
period of the order of\ the inverse time delay between pulses $T^{-1}$ [see
Ramsey phases $\phi_{R}$ (\ref{51c}) and (\ref{56d}) for single- and
double-loop interferometers]. Evidently, this new type of Ramsey fringes
arises because even for Raman pulses with copropagating effective wave
vectors, in the condition of the Doppler phase cancellation, time separations
between pulses $T_{j}$ enter with weight factors $\mathbf{k}_{j},$ while these
factors are absent in expressions for the Ramsey phase. In contrast to the
previously proposed \cite{c20} and observed \cite{c6} Raman-Ramsey fringes on
Doppler broadened two-quantum transitions based on the use of Raman pulses
having counterpropagating effective wave vectors, Ramsey fringes found here do
not undergo recoil splitting, i.e., they are centered at the two-quantum line center.

Double-loop interferometer shown in Fig. \ref{f1}(c) allows one to get the
phase where all cubic in $T$ terms are eliminated and can serve as\ an
accelerometer. We expect that owing to this cancellation, the accuracy of the
acceleration measurement can be increased.

Thus in this paper we propose the following scenario for the atomic gravity
and rotation sensing. First one measures acceleration $\mathbf{a}$ using three
mutually perpendicular double-loop accelerometers. Afterwards, one uses the
spatial-domain atomic gyroscope \cite{c4}, orients it in the plane
perpendicular to $\mathbf{a}$ and measures the component of the rotation
frequency $\boldsymbol{\mathbf{\Omega}}$ along $\mathbf{a}.$ Then one uses two
mutually perpendicular time-domain double-loop gyroscopes to get components of
$\boldsymbol{\mathbf{\Omega}}$ perpendicular to $\mathbf{a.}$

Further obvious step here could be the consideration of multiple-loop
interferometers. In general for an interferometer consisting of $\ell$
unidirectional pulses (first and last are $\pi/2$ pulses, others are $\pi$
pulses), having the same wave vector, one has $\ell-2$ parameters $t_{j}%
=T_{j}/T_{\ell}$ $\left(  j=2,~\ldots,~\ell-1\right)  $, which can be chosen
as a root of the system of $\ell-2$ equations%
\begin{equation}
1-2t_{\ell-1}^{u_{j}}+2t_{\ell-2}^{u_{j}}+\ldots+\left(  -1\right)  ^{\ell
}2t_{2}^{u_{j}}=0\label{77}%
\end{equation}
to eliminate phase terms evolving as $T_{\ell}^{u_{1}},$ \ldots,$T_{\ell
}^{u_{\ell-2}}.$ In the array $\mathbf{u}=\left(  u_{1},\ldots,u_{\ell
-2}\right)  $ one of parameters $u_{j}$ has to be equal to 1, to
eliminate\ the Doppler phase, while the choice of other parameters depends on
phenomena one would like to sense using the interferometer. Remaining
noneliminated phase terms evolving as $T_{\ell}^{v}$ can be obtained from
Table \ref{t1} by replacement of%
\begin{equation}
T_{3}^{v}\rightarrow\dfrac{1-2t_{\ell-1}^{v}+2t_{\ell-2}^{v}+\ldots+\left(
-1\right)  ^{\ell}2t_{2}^{v}}{1-2^{1-v}}T_{\ell}^{v}.\label{78}%
\end{equation}
The case of four-pulse interferometers is considered above in detail. Using
five-pulse interferometers one can eliminate $T_{5}^{4}$ dependencies. For
gyroscope $\mathbf{u}=\left(  1,2,4\right)  ,$ we found the numeric solution
of system (\ref{77}),%
\begin{equation}
t_{2}=0.16615,~t_{3}=0.54057,~t_{4}=0.87442.\label{79}%
\end{equation}
This gyroscope phase is given by%
\begin{equation}
\phi_{m}=-1.0143\times10^{-2}\boldsymbol{\mathbf{\Omega}}\cdot\left(
\mathbf{k}\times\mathbf{a}\right)  T_{5}^{3}+O\left(  T_{5}^{5}\right)
.\label{80}%
\end{equation}
For the Cs gyroscope with the same parameters as in Table \ref{t2}, $\phi
_{m}\approx-41.2.$

For the five-pulse accelerometer, $\mathbf{u}=\left(  1,3,4\right)  ,$ one
finds%
\begin{equation}
t_{2}=0.20594,~t_{3}=0.594\,37,~t_{4}=0.88843,\label{81}%
\end{equation}
and the phase is given by%
\begin{equation}
\phi_{m}=2.1557\times10^{-2}\mathbf{k}\cdot\mathbf{a}T_{5}^{2}+O\left(
T_{5}^{5}\right)  .\label{82}%
\end{equation}
For the Cs accelerometer with the same parameters as in Table \ref{t3},
$\phi_{m}\approx-2.00\times10^{6}.$

Another example of five-pulse interferometer is a \textquotedblleft figure 8
1/2\textquotedblright\ interferometer proposed in \cite{c121} to eliminate
$T_{5}^{2}$ and $T_{5}^{3}$ terms, such as the dominate contribution to the
phase is caused by the influence of space-time curvature.

Consider also six-pulse interferometer. One can use this pulse sequence to
eliminate all phase terms arising as a result of the atom motion to the
accuracy $T_{6}^{4}$ choosing $\mathbf{u}=\left(  1,2,3,4\right)  .$ The
numerical solution of the system (\ref{77}) is given by
\begin{equation}
t_{2}=9.5492\times10^{-2},~t_{3}=0.34549,~t_{4}=0.65451,~t_{5}%
=0.90451.\label{83}%
\end{equation}
Of course, one cannot use this interferometer to sense atomic motion. But one
can use it for the precise measurement of the other spectroscopic data. The
accuracy of this measurement improves when the time of evolution $T_{6}$
increases. Earth gravity and rotation superimpose limitation on this time.
This time can be increased in the microgravity environment. We actually
propose to use the six-pulse interferometer instead of transferring
experiments into microgravity environment.

Possible examples here are measurement of levels polarizability \cite{c21} or
observation of the Aharonov-Bohm effect \cite{c22}. We also believe that one
can insert additional $\pi$ pulses in the scheme of the recoil frequency
measurement \cite{c6} to make this measurement independent on the gravity,
gravity gradient and the Earth rotation.

\begin{figure} [htbp]
\vspace{0.5cm}
\includegraphics[height=12.0cm,width=12.0cm]{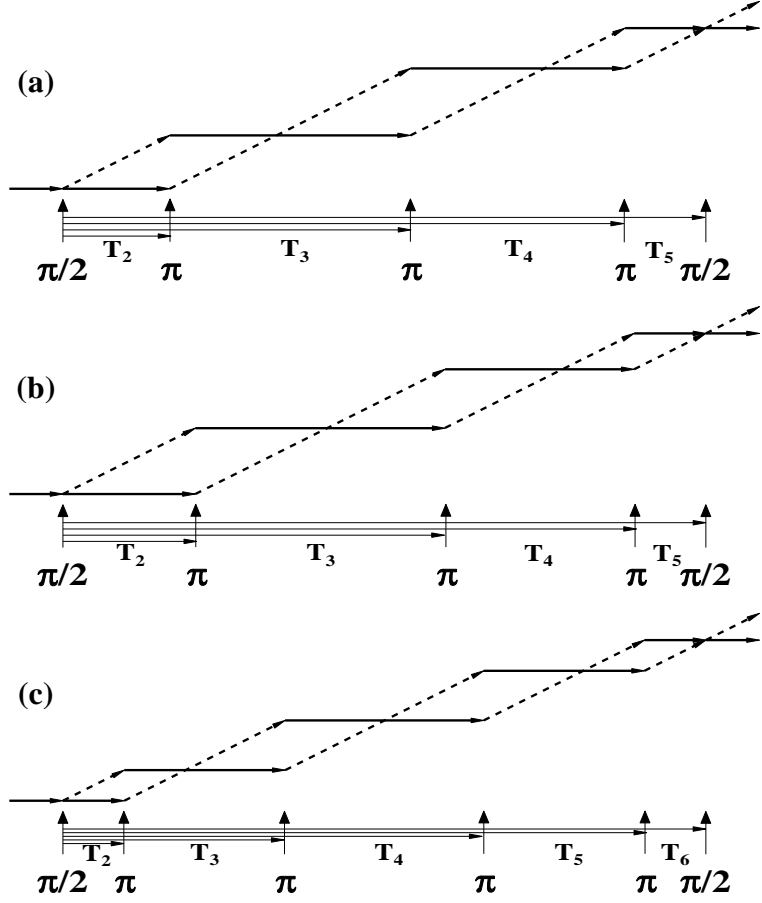}
\caption{Recoil diagrams for the five-pulse gyroscope (a), accelerometer (b)
and six-pulse interferometer insensitive to both rotation and acceleration.}
\label{f4}
\end{figure}

Recoil diagrams for five- and six-pulse interferometers are shown in Fig.
\ref{f4}.
Regarding the six-pulse interferometer, one sees that it has to be protected
from two kinds of the stimulated echo processes produced by the sequences of
pulses $\left\{  1,2,5,6\right\}  $ and $\left\{  1,3,4,6\right\}  .$ A
similar note can be made regarding the \textquotedblleft figure 8
1/2\textquotedblright\ interferometer \cite{c121}, where both stimulated echo
produced by pulses $\left\{  1,2,4,5\right\}  $ and one-loop interferometer
produced by pulses $\left\{  1,3,5\right\}  $ have to be excluded. An
effective technique of eliminating all these unwanted signals is using Raman
pulses with nonequal effective wave vectors, as we demonstrated and analyzed
above for the double-loop atomic gyroscope.

\acknowledgments

The authors are grateful to P. R. Berman for fruitful discussions and K.-P.
Marzlin for numerous comments. This work was supported by DARPA.


\begin{thebibliography}{99}
\bibitem {c151}Density matrix in the Wigner representation could have negative
sum of the diagonal matrix elements, $F\left(  \mathbf{r,p,}t\right)  \equiv
Tr\left[  \rho\left(  \mathbf{r,p,}t\right)  \right]  .$ Moreover, from Eqs.
(\ref{49}) and (\ref{55}) and corresponding expressions for $\rho_{gg}\left(
\mathbf{r,p,}t\right)  ,$ one concludes that even in the absence of the
forces, for the large recoil momentum, $\hbar k_{1}\gg\Delta p,$ $F\left(
\mathbf{r,p,}t\right)  $ can be negative in the vicinities of the points
$\mathbf{p=}\pm\hbar\mathbf{k}_{1}/2.$ This circumstance should not create any
difficulty in the interpretation, because it is impossible to introduce
distribution function over positions and momenta for the gas of atoms with
quantized center-of-mass degrees of freedom, and function $F\left(
\mathbf{r,p,}t\right)  $ does not have the meaning of this distribution function.
\end{thebibliography}
\end{document}